\documentclass[journal ]{new-aiaa}
\usepackage[utf8]{inputenc}
\usepackage{textcomp}

\usepackage{graphicx}
\usepackage{amsmath, bm}
\usepackage[version=4]{mhchem}
\usepackage{siunitx}
\usepackage{longtable,tabularx}
\usepackage{subfigure}

\DeclareGraphicsExtensions{.pdf,.png,.jpg,.svg,.eps}
\title{Real-Time Prediction of Probabilistic Crack Growth with a Helicopter Component Digital Twin}

\author{Xuan Zhou\footnote{PhD student, School of Aeronautic Science and Engineering.}, Shuangxin He\footnote{PhD student, School of Aeronautic Science and Engineering; Corresponding author: heshuangxin@buaa.edu.cn}, Leiting Dong\footnote{Professor, School of Aeronautic Science and Engineering; Corresponding author: ltdong@buaa.edu.cn.}}
\affil{Beihang University, Beijing, China, 100191}

\author{Satya N. Atluri\footnote{Professor and Presidential Chair, Department of Mechanical Engineering; Fellow AIAA.}}
\affil{Texas Tech University, Lubbock, Texas, USA, 79409}

\begin{document}

\maketitle

\begin{abstract}
    To deploy the airframe digital twin or to conduct probabilistic evaluations of the remaining life of a structural component, a (near) real-time crack-growth simulation method is critical. In this paper, a reduced-order simulation approach is developed to achieve this goal by leveraging two methods. On the one hand, the symmetric Galerkin boundary element method–finite element method (SGBEM-FEM) coupling method is combined with parametric modeling to generate the database of computed stress intensity factors for cracks with various sizes/shapes in a complex structural component, by which hundreds of samples are automatically simulated within a day. On the other hand, machine learning methods are applied to establish the relation between crack sizes/shapes and crack-front stress intensity factors. By combining the reduced-order computational model with load inputs and fatigue growth laws, a real-time prediction of probabilistic crack growth in complex structures with minimum computational burden is realized. In an example of a round-robin helicopter component, even though the fatigue crack growth is simulated cycle by cycle, the simulation is faster than real-time (as compared with the physical test). The proposed approach is a key simulation technology toward realizing the digital twin of complex structures, which further requires fusion of model predictions with flight/inspection/monitoring data.
\end{abstract}

\section*{Nomenclature}

{\renewcommand\arraystretch{1.0}
    \noindent\begin{longtable*}{@{}l @{\quad=\quad} l@{}}
        $a$ &  crack length \\
        $C, \gamma$ & generalized Frost–Dugdale law coefficients \\
        $\mathcal{D}$ & database of fracture mechanics simulation \\
        $K$& Stress Intensity Factor \\
        $K^{b}$ & Stress Intensity Factor under a benchmark load \\
        $L_{a}, L_{c}$ & initial crack shape parameters \\
        $P^b$  & benchmark load \\ 
        $ r $ & threshold of the cumulative variance contribution rate of principal components \\ 
        $S$   & spline parameter set of the original front curve nodes \\
        $s$   & spline parameter set of the original nodes \\
        $tp$   & trim percentage of a curve \\
        $R$  & stress ratio \\
        $x$  & equispaced nodes along a crack front\\
        $X$  & non-equispaced nodes along a crack front\\
        
        $\beta$ & cumulative variance contribution rate of principal components\\
        $\Delta K $ & Stress Intensity Factor range\\
        $\varepsilon$  & training or test error \\
        \multicolumn{2}{@{}l}{Subscripts}\\
        equi & processed form with equispaced nodes \\
        ori & original form with non-equispaced nodes \\
    \end{longtable*}}

\section{Introduction}

\lettrine{H}{elicopter} dynamic components are some of the most fatigue-critical components on the helicopter structure, which consist of main and tail rotors, rotor hub, drive shafts and gear boxes, and whose integrity during service life directly affects the safety of the helicopter. Different from fixed-wing aircraft, helicopter dynamic components are mostly subjected to low amplitude and high-frequency loads. Thus, the duration in which a small crack propagates to the critical size, which leads to failure, is relatively short. Therefore, helicopters are more likely to incur accidents caused by fatigue damage.

For a long time, the safe life method has been used to address the structural fatigue of helicopter dynamic components. In the safe life method, all key fatigue components, assumed to have no preexisting defects, are assigned a finite service life, which is determined by the Palmgren Miner nominal stress rule \cite{minerCumulativeDamageFatigue1945} or other rules, and the components are required to be retired when reaching the service life. However, fatigue failures are often caused by unexpected factors such as process defects, maintenance errors and damage induced by usage, which are not considered in the safe life method and result in the occurrence of numerous flight accidents, indicating that the safe life design cannot completely guarantee the safety of helicopter structures.

In the past few decades, there was a growing interest in using Damage Tolerance Methodology to determine the life and inspection schedules of helicopter components. Damage tolerance method relies on the assumption that inevitably there are initial damages which will subsequently grow over a period of time prior to the catastrophic failure \cite{zhaoChallengesDamageTolerance2011}. Fracture mechanics methodology is used to predict the fatigue crack growth in the structure, and inspection intervals are devised to ensure that the crack does not grow to the critical size. However, there are many uncertainties in the crack growth process. For example, loads, structural geometry, and crack growth behaviors may have some level of dispersions between various aircraft and flight missions. As a result, methods are proposed to track and control such uncertainties within the life of structural components.

Individual structural life and maintenance managements are attracting increasing interests recently, where Digital Twin is a potential solution. The digital twin philosophy, originally presented by Michael Grieves in 2003 \cite{grievesDigitalTwinManufacturing2014}, aims to create a digital representation of the corresponding physical product in the life-cycle of the product. In 2009, the National Research Council Canada proposed a roadmap for the probabilistic life usage monitoring of helicopters \cite{cheungRoadmapHelicopterLife2009} which can be seen as a preliminary version of the digital twin concept for helicopters. In 2011, the U.S. Air Force Research Laboratory (AFRL) proposed to apply the digital twin philosophy to the life management of airframes \cite{tuegelReengineeringAircraftStructural2011}.NASA Langley Research Center \cite{leserDigitalTwinFeasibility2020} conducted a feasibility study in which in-situ diagnostics and prognostics are coupled in a probabilistic framework to track and control uncertainties in structural fatigue. National Research Council Canada \cite{renaudDemonstrationAirframeDigital2020,liaoAirframeDigitalTwin2020} also proposed to use a CF–188 full-scale component test to assess the adaptability of ADT approach for Royal Canadian Air Force (RCAF) fleets.

In order to deploy the airframe Digital Twin or to conduct individualized evaluations of the remaining life of a structural component, a (near) real-time crack growth simulation method is critical. However, full-order fracture mechanics simulation methods cannot fulfill this requirement. T. D. West at Arnold Air Force \cite{westDigitalThreadDigital2017} presented concerns about the cost of developing and deploying an airframe digital twin. One of his examples is that a finite element simulation of the crack growth in a plate with notches and holes (modeled with 5.5 million DoFs finite elements) required a 4-days wall-clock computational time \cite{cerroneEffectsModelingAsManufactured2014}. In this regard, the reduced-order model (ROM), which approximates the full-order model and significantly reduces the computational burden with an acceptable accuracy, is a feasible alternative.

A limited set of investigations \cite{sankararamanUncertaintyQuantificationFatigue2011,hombalSurrogateModeling3D2013,leserProbabilisticFatigueDamage2017,keprateAdaptiveGaussianProcess2017,keprateSurrogateModelPredicting2017} have been reported on using ROMs for fracture mechanics simulations, which are focused on establishing the relation between crack sizes/shapes to crack front Stress Intensity Factors. In general, the reduced order modelling of the crack growth can be divided into two stages: the offline stage and the online stage. In the offline stage, a database containing various cracks is established by full-order fracture mechanics simulations, and the reduced-order model is constructed using the database; while in the online stage, the ROM is used in the crack growth prediction and the remaining life assessment as a surrogate of the full-order simulation. During this procedure, constructing the database of fracture mechanics simulations is crucial, which nevertheless requires significant computational burden, especially when cracks in complex three-dimensional structural components are considered. In \cite{liDynamicBayesianNetwork2017}, analytical solutions are used to calculate the Stress Intensity Factors of simple-shape cracks. In \cite{leserDigitalTwinFeasibility2020} a characteristic crack length was defined and related to crack-front SIFs, where influences of detailed crack-shapes on SIFs are disregarded. Other studies limit the database of cracks by merely sampling from a sequence of cracks in a crack growth simulation \cite{hombalSurrogateModeling3D2013,leserProbabilisticFatigueDamage2017,leserDigitalTwinFeasibility2020}. With this in mind, a simple, efficient and automatic approach which can quickly generate models of various cracks and efficiently perform fracture mechanics simulations, is desirable for constructing the ROM, and for realizing the digital twin of complex structural components.

In this paper, a reduced-order simulation approach is developed to address this challenge by leveraging two methods. Firstly, a fast and efficient fracture mechanics simulation method, the SGBEM (Symmetric Galerkin Boundary Element Method) - FEM (Finite Element Method) coupling method, is adopted to generate the fracture mechanics database of cracks with various sizes and shapes in a complex structural component, which is the latest development within a series of fracture mechanics simulation methods. The basic thought of this series of works is to model the complex structure with an unchanging finite element model with no cracks, and to model cracks in an efficient way, either by complex variable, special functions, or by SGBEMs. The original version is the Finite Element Alternating method (FEAM) \cite{nishiokaAlternatingMethodAnalysis1983}, which uses the Schwartz-Neumann alternation between the finite element solution of the entire structure and the analytical solution for an infinite body containing the crack. It was later successfully extended to BIE (Boundary Integral Equation) -FEM alternating method to simulate arbitrarily 2D curved cracks in \cite{parkMixedModeFatigue1998}, and SGBEM-FEM Alternating method for arbitrary 3D non-planar embedded and surface cracks in \cite{nikishkovSGBEMFEMAlternatingMethod2001,hanSGBEMCrackedLocal2002}. Recent developments are the SGBEM super element - FEM coupling method \cite{dongSGBEMUsingNonhypersingular2012} in which a SGBEM "Super Element" describing a local subdomain containing arbitrary cracks is directly coupled with finite elements, using the simple assembly procedure. It is shown in the examples of this study that, with the SGBEM super element - FEM coupling method, hundreds of samples of cracks can be simulated in one day using an ordinary desktop PC. Furthermore, the reduced-order model, which is composed of PCA-based dimensionality reduction of crack front coordinates and a surrogate-based Stress Intensity Factor computation, is established. Several regression methods are explored to establish the surrogate model and their performance is compared. Finally, the proposed method is applied to the crack growth prediction and the remaining life evaluation of a round- robin test helicopter component. Even though the crack-growth is simulated cycle by cycle, the total computational time is only 20 minutes using an ordinary desktop PC. When the step size (load cycles per calculation) is increased, near real-time simulations can be realized without compromising the simulation accuracy.

The remainder of this paper is organized as follows. Section \ref{section: database} introduces the method of constructing a fracture mechanics database based on the "SGBEM super element - FEM coupling method". Section \ref{section: ROM} presents a data-driven reduced-order modeling approach using the constructed database, and further compares the Stress Intensity Factor predictions using different regression methods. In Section \ref{section: application}, an example of a round-robin helicopter component is used to demonstrate the capability of the constructed reduced-order model for deterministic crack growth predictions and probabilistic remaining life evaluations. In Section \ref{section: conclusion}, this study is completed with some concluding remarks.

\section{ An Automatic and Efficient Fracture Mechanics Simulation Approach to Generate a Database of SIFs for Various Cracks}
\label{section: database}
To construct the database of SIFs for various cracks, there are generally three steps: (1) Characterize cracks of interests with several parameters, and sample in the parameter space to generate cracks of various locations/shapes/sizes. (2) For each sample crack, construct the CAE model with meshes, material parameters, loads, etc. (3) Conduct batch fracture mechanics simulations to obtain the Stress Intensity Factors for each sample crack. In this process, the human labor of mesh-generation in step (2) and the computational burden to simulate the structure with various cracks, can be significantly reduced by the exploring the advantages of the "SGBEM super element - FEM coupling method" \cite{dongSGBEMUsingNonhypersingular2012}.

In the SGEBM super element - FEM coupling method \cite{dongSGBEMUsingNonhypersingular2012}, the uncracked global structure is modelled by the finite element method (FEM), while the three-dimensional crack with a local finite-sized subdomain containing the crack is modelled by a SGBEM super element. The boundary mesh for the SGBEM cracked subdomain can be automatically generated using a simple crack-surface mesh and the FEM mesh of the uncracked structure, by grouping FEM elements in a finite region surrounding the crack and taking only the information of the boundary/surface elements. A FEM-type of stiffness matrix of the SGBEM super element containing the crack can be obtained by rearranging the boundary integral equations \cite{dongSGBEMUsingNonhypersingular2012}, therefore the FEM model of the uncracked structure and the SGBEM model of the cracked subdomain can be directly coupled by the assembly of stiffness matrices. 

\begin{figure}[hbt!]
    \centering
    \includegraphics[width=.8\textwidth]{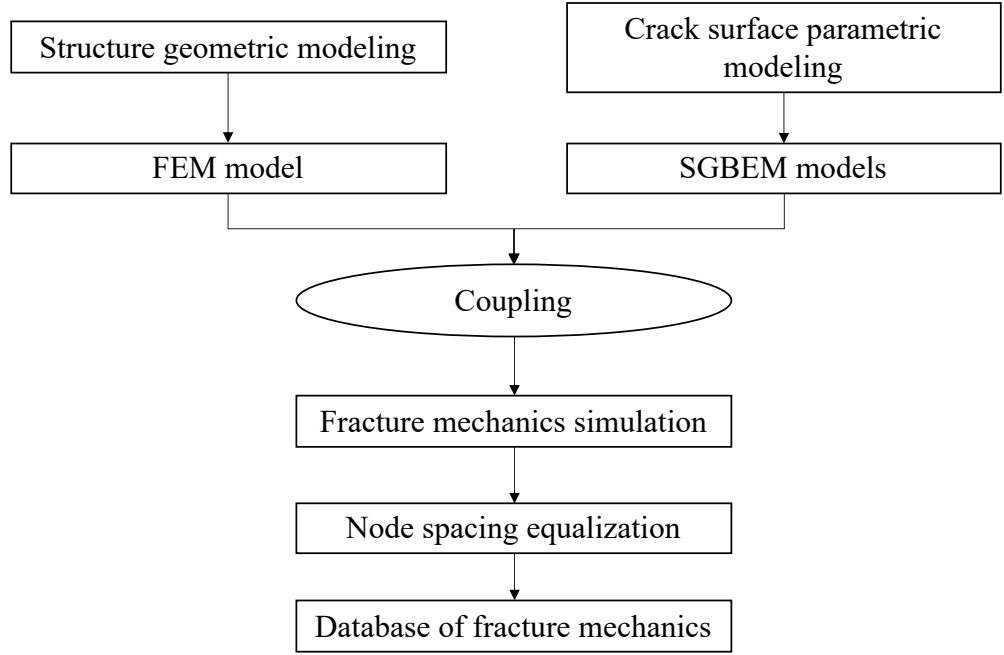}
    \caption{Flow chart of the construction of the Database of SIFs for Various Cracks}
    \label{fig: database_flowchart}
\end{figure}

Such an approach of modeling the uncracked structure (by FEM) and the cracked local sub-domain (by SGBEM) with different methods can significantly simplify and accelerate the process of constructing the database of SIFs for various cracks, as shown in Fig. \ref{fig: database_flowchart}. On the one hand, because of the independence of the FEM mesh and crack surface mesh in the preprocessing stage, step (2) can be significantly simplified. It is worth noticing that the finite element model of the uncracked global structure needs to be constructed only once, and the surface mesh of crack with various shapes/sizes can be automatically done by parametrized modeling tools, as shown in the next subsection. On the other hand, due to the accuracy of SGBEM for modeling a small cracked region and the efficiency of FEM for modeling uncracked complex structure, such a combined simulation approach can significantly reduce the computational time in step (3). The computational accuracy and simulation efficiency of the coupled SGBEM-FEM method for fracture and fatigue analyses of 2D and 3D complex structures have also been systematically evaluated and demonstrated in \cite{dongFractureFatigueAnalyses2013a, dongFractureFatigueAnalyses2013}. In this paper, we demonstrate the procedure of constructing the database of various cracks by this approach using a round-robin helicopter component, as shown in the next 2 subsections.

\subsection{The Round-robin Helicopter Component and the FEM Model}
\label{section: structure_case}

A round-robin helicopter component, the lift frame described in Cansdale and Perret \cite{cansdaleHelicopterDamageTolerance2002} and Irving, Lin and Bristow \cite{irvingDamageToleranceHelicopters2003}, is used for demonstration. 

\subsubsection{Geometric and Finite Element Modeling}

\begin{figure}[hbt!]
    \centering
    \includegraphics[width=.7\textwidth]{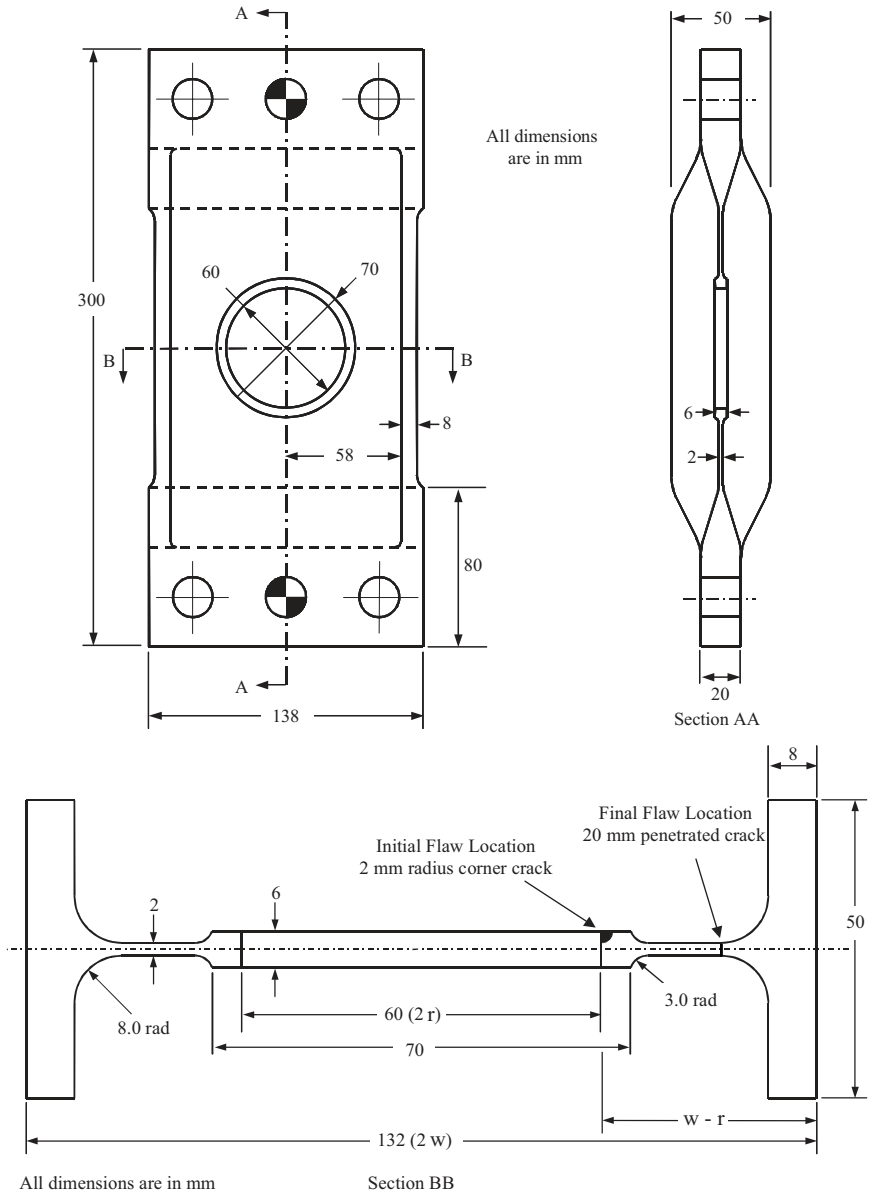}
    \caption{Geometry of the helicopter lift frame. Adapted with permission from \cite{newmanCrackGrowthPredictions2006}. Copyright 2006 John Wiley and Sons.}
    \label{fig_geometry}
\end{figure}

The geometry of the component is a flanged plate with a central lightning hole, where a 2 mm corner crack is initiated, (see Fig. \ref{fig_geometry}) according to which the finite element model was constructed. Then the geometry model is constructed in SOLIDWORKS and  imported in MSC/PATRAN to generate the finite element model of the uncracked structure, where uniform tensile loads is applied at both ends, and the materials and load spectrum are given as follows.

\subsubsection{Material}

As shown in \cite{tiongDamageToleranceAnalysis2009}, the helicopter component was made of AL 7010-T73651, whose average room temperature fracture toughness in the (L–T) orientation is 33.4 $\rm MPa\sqrt{m}$. It is assumed to have a Young’s modulus of 70,000 MPa and a Poisson’s ratio of 0.3 in this study.

\subsubsection{Load Spectrum}

\begin{figure}[hbt!]
    \centering
    \includegraphics[width=.8\textwidth]{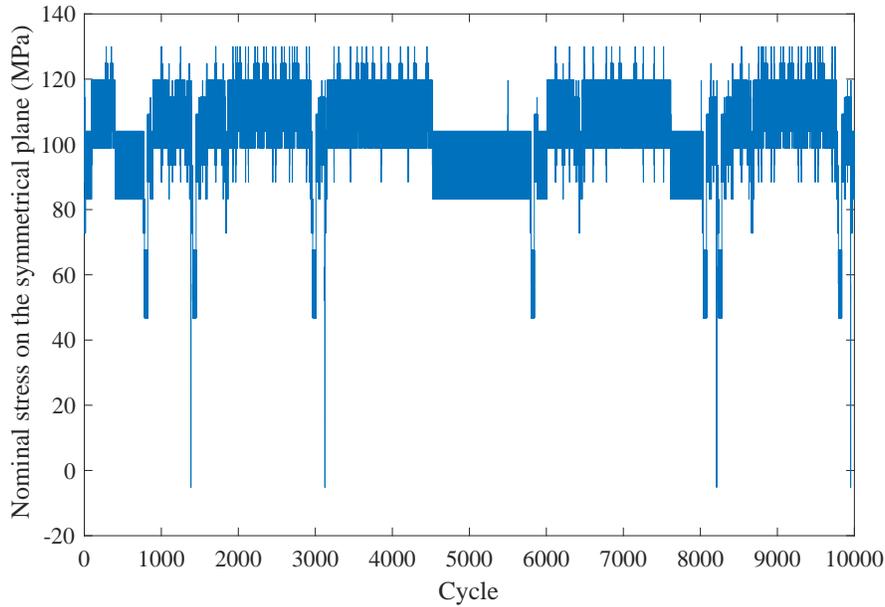}
    \caption{A sample of the Asterix spectrum.}
    \label{fig_asterix}
\end{figure}

The component was subjected to the Asterix spectrum, a representative helicopter spectrum loading, which is shown in Fig. \ref{fig_asterix}. The Asterix spectrum was derived from the strain data measured on a helicopter lift frame. It has an average R value of 0.82 and a limited number of negative R ratio excursions \cite{vaughanLifePredictionHigh2003}. There are very few load-cycles with an R ratio of less than R = 0.7. The Asterix spectrum, which represents a single load block, corresponds to 190.5 flight hours with 371,610 load cycles. In this paper, the largest far field stress in the spectrum is set to 130 MPa.

\subsection{Constructing Sample of Various Cracks based on Parametric Modeling}

The SGBEM super element - FEM coupling method takes two inputs for the fracture mechanics simulation, the finite element model of the uncracked structure and the crack surface mesh. In order to construct the database of SIFs with various cracks, a large number of cracks with various sizes/shapes need to be generated and meshed, which can be time-consuming and labor-intensive if done manually. In this paper, a PATRAN-based parametric modeling tool is implemented to automate and speed up the process.

\begin{figure}[htbp]
    \centering
    \includegraphics[width=.9\textwidth]{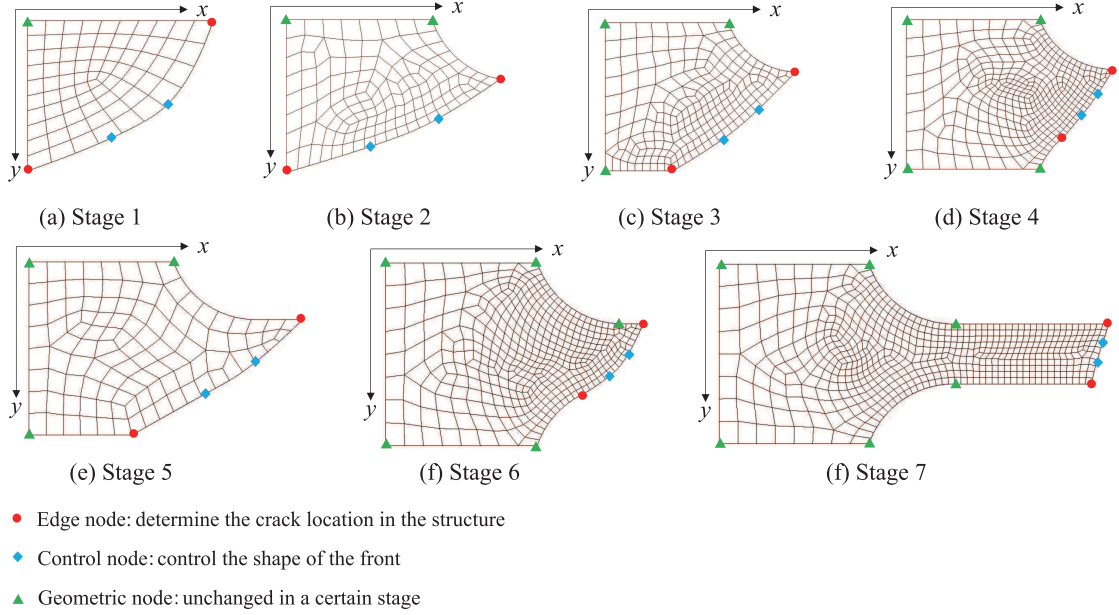}
    \caption{Typical crack surface geometries and meshes at various stages.}
    \label{fig: crack surface}
\end{figure}

The process of crack growth in the helicopter component can be divided into several stages as shown in Fig. \ref{fig: crack surface}. At each stage, the PATRAN parameterized modeling script is generated by an in-house MATLAB code, which is then run to automatically to generate a meshed crack surface. Such a parametrized modeling process is as follows.

\begin{enumerate}
    \item In each stage, divide the boundaries of the crack surface into geometric boundaries which can be determined from unchanged geometric nodes (marked as green triangles), and the crack front which can be determined by B-spline interpolation of edge nodes (red circles) and control nodes (blue diamond), as shown in Fig. \ref{fig: crack surface}. Then, generate a sufficient number of edge and control nodes by Latin hypercube sampling method. 
    \item Using the unchanged geometric nodes and the generated edge and control nodes, generate a PATRAN parametric modeling script and define the crack surface in the order of points, curves and the surface.
    \item Generate the crack-surface mesh automatically by the PATRAN mesh generation module.
\end{enumerate}

In the Latin hypercube sampling, there are six sampling parameters at each stage, which is divided into two types. The first type is the coordinate parameters of the two edge nodes (in a local $x-y$ plane, $x$ coordinates for the line on the top or the line on the bottom, $y$ coordinates for the line on the left, and the trim percentage of the curve ($tp$)), by which the location of the crack can be determined; The second type is the disturbance coefficients $\epsilon \in [0,1] $ which determine the shape of the crack front. The original location of the control nodes are the two trisection points of the line segment between the upper and lower edge nodes, which are then random disturbed in two directions. The amplitude of the disturbance is set as the disturbance coefficient multiplied by 1/3 of the distance between the upper and lower nodes in coordinate $x$ or $y$. The sampling parameters of each stage are shown in the Table. \ref{tab:sampling_parameters}

\begin{table}[htbp]
    \centering
    \caption{Sampling parameters of Latin hypercube sampling at each stage}
    \begin{tabular}{ccccccccc}
    \hline
    parameter & \multicolumn{2}{c}{coordinates 1} & \multicolumn{2}{c}{coordinates 2} & \multicolumn{4}{c}{disturbance coefficient} \\
    stage & type  & range & type  & range & $x_{1/3}$ & $y_{1/3}$ & $x_{2/3}$ & $y_{2/3}$ \\
    \hline
    1     & $x$     & [2,5] & $y$     & [2,6] & [0,1] & [0,1] & [0,1] & [0,1] \\
    2     & $tp$    & [0,1] & $y$     & [2,6] & [0,1] & [0,1] & [0,1] & [0,1] \\
    3     & $tp$    & [0,1] & $x$     & [0,5] & [0,1] & [0,1] & [0,1] & [0,1] \\
    4     & $tp$    & [0,1] & $tp$    & [0,1] & [0,1] & [0,1] & [0,1] & [0,1] \\
    5     & $x$     & [7.828,12] & $x$     & [0,5] & [0,1] & [0,1] & [0,1] & [0,1] \\
    6     & $x$     & [7.828,12] & $tp$    & [0,1] & [0,1] & [0,1] & [0,1] & [0,1] \\
    7     & $x$     & [7.828,20] & $x$     & [7.828,20] & [0,1] & [0,1] & [0,1] & [0,1] \\
    \hline
    \end{tabular}%
    \label{tab:sampling_parameters}%
\end{table}%

\subsection{Batch SGBEM-FEM Simulation to Calculate the Stress Intensity Factor}

\begin{figure}[hbt!]
    \centering
    \includegraphics[width=.8\textwidth]{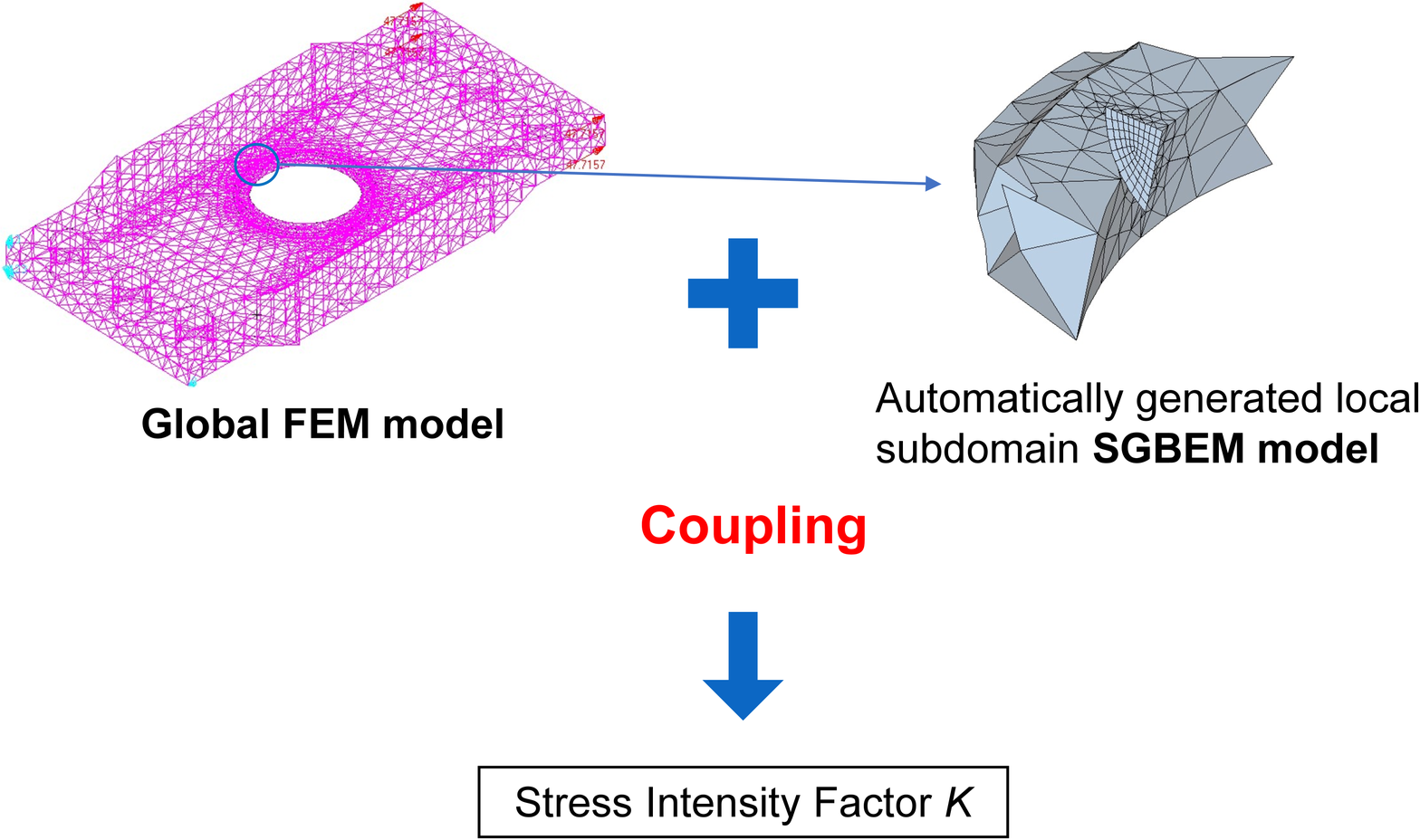}
    \caption{Schematic diagram of the SGBEM super element - FEM coupling method for the fracture mechanics simulation.}
    \label{fig: sgbem-fem}
\end{figure}

\begin{figure}[hbt!]
    \centering
    \includegraphics[width=.5\textwidth]{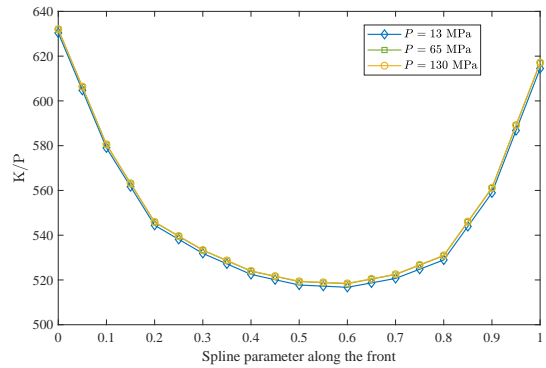}
    \caption{For a 1/4 corner crack with a radius of 2.5 mm, the ratio of the Stress Intensity Factor to the load amplitude $K/F$ remains unchanged under three different load amplitudes.}
    \label{fig: sif_diff_loads}
\end{figure}

As shown in Fig. \ref{fig: sgbem-fem}, after the construction of the FEM model and a sufficient number of crack surface samples, the in-house code of the SGBEM super element - FEM coupling method is called in batches to calculate the Stress Intensity Factor of each sample crack under a benchmark load. The reason why a benchmark load is used is that the Stress Intensity Factor is proportional to the load, as shown in Fig. \ref{fig: sif_diff_loads}. It is worth noting that, due to the high efficiency of the SGBEM super element - FEM coupling method, about three hundred samples in the first stage were simulated in half a day using an ordinary desktop (Intel Core i7-6700, 16 GB memory).

%

The generated database is denoted as $ \mathcal{D}_{ori}=\{ (\mathbf{C}_{ori,i},K_{ori,i}^{b}):i=1,2,...,N)\}$, where $\mathbf{C}_{ori,i}$ is the location vector of the $i^{th}$ sample and $K_{ori,i}^{b}$ is the corresponding SIF under a benchmark load.

The quality of the database is crucial for constructing ROM, which in this case is the accuracy of the computed Stress Intensity Factors. The computational accuracy of the SGBEM-FEM method has been systematically verified in our two-part paper in 2013 \cite{dongFractureFatigueAnalyses2013,dongFractureFatigueAnalyses2013a}, where many examples of 2D through-thickness cracks and 3D embedded/surface cracks. 

\subsection{Nodal Spacing Equalization}

In the parametric crack-surface modeling process, the generated number of nodes at the crack front may be different for various cracks, and the nodal spacing may also be non-uniform. Therefore, to establish the relation between nodal coordinates and the SIFs at crack fronts for different cracks, a nodal spacing equalization procedure in \cite{hombalSurrogateModeling3D2013} is used here to set the same number of equidistant nodes along the crack front for a consistently parameterized representation of the crack fronts between different cracks.

The node spacing equalization aims to set the same number of equidistant nodes on the front curve to generate a consistently parameterized representation of the crack fronts between different cracks. In this paper, the method in \cite{hombalSurrogateModeling3D2013} is used, but corresponding improvements are made for specific problems.

For a crack front curve with non-equispaced nodes, we can represent the nodes with coordinates:

\begin{equation}
    \mathbf{X}=\{\mathbf{X}_{1},\mathbf{X}_{2},...,\mathbf{X}_{m}\}
\end{equation}
where $m$ is the number of the nodes. $\mathbf{X}_{i}$ is the location vector for the $ i^{th}$ node.

A cubic spline interpolation is used to generate a spline representation $f_1(\xi)$ of the crack front, where $ \xi \in [0,1]$ is the spline’s natural parameter. Assuming that $p+1$ equispaced nodes along the crack front are to be generated, the spline parameter for these nodes are:

\begin{equation}
    \mathbf{s}=\{0,\frac{1}{p},\frac{2}{p},...,\frac{p-1}{p},1\}
\end{equation}

and the location vector of the generated equispaced nodes along the crack-front can be calculated by $ \mathbf{x}=f_1(\xi)$:

\begin{equation}
    \mathbf{x}=\{\mathbf{x}_{0},\mathbf{x}_{2},...,\mathbf{x}_{p}\}
\end{equation}

For each node $\mathbf{x}_{i}$, we can have two representations: Cartesian coordinate and corresponding spline parameter.
 
Meanwhile, the spline parameter values corresponding to the original non-equispaced nodes $\mathbf{X}_{i}$ can be obtained by $\xi=f_1^{-1}(\mathbf{X})$, which can be expressed as:

\begin{equation}
    \mathbf{S}= \{0,\xi_{1},\xi_{2},...,\xi_{m-1},1\}
\end{equation}

The cubic spline interpolation is also used to fit the Stress Intensity Factor.

\begin{equation}
    \mathbf{K}^{b}=f_{2}(\xi)
\end{equation}

Then the corresponding Stress Intensity Factor at the newly generated equispaced nodes $\mathbf{K}_{equi}^{b}$ can be obtained by the spline function $f_{2}$.

\begin{equation}
    \mathbf{K}_{equi}^{b}=\{K_{0},K_{1},K_{2},...,K_{p}\}
\end{equation}

\begin{figure}[hbt!]
    \centering
    \includegraphics[width=.7\textwidth]{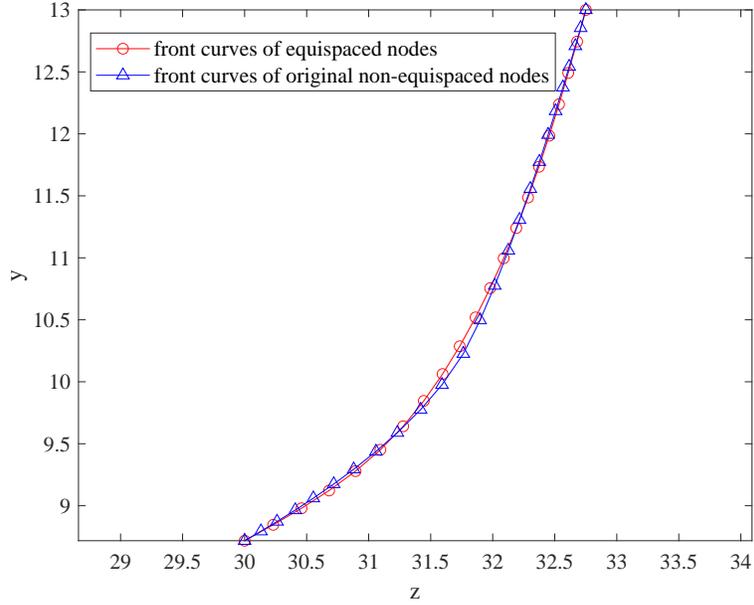}
    \caption{	Comparison of the generated equispaced nodes and the original non-equispaced nodes of a sample crack.}
    \label{fig: front_curve_cmp}
\end{figure}

A comparison of the generated equispaced nodes and the original non-equispaced nodes of a sample crack is shown in Fig. \ref{fig: front_curve_cmp}.

By equispacing all samples, a set of crack front curves with a consistently parameterized representation can be obtained, which can be represented as a matrix $\mathbf{C}_{equi}$. Then the database with equispaced nodes is denoted as $ \mathcal{D}_{equi}=\{(\mathbf{C}_{equi,i},K_{equi,i}^{b}):i=1,2,...,N)\}$.

\section{Data-driven Reduced-order Fracture Mechanics Modeling of Cracked Complex Structure}
\label{section: ROM}

In this paper, a reduced-order model $\mathcal{M}^{r}$ is used to replace the time-consuming fracture mechanics simulation model $\mathcal{M}$. In this model, the input consists of parameterized parameters of the crack-front, and the output is the corresponding Stress Intensity Factors along the crack front under a benchmark load $\mathbf{P}^{b}$. The ROM can be expressed as:

\begin{equation}
    \mathbf{K}^{b}=\mathcal{M}^{r}(\mathbf{a})
\end{equation}
where $\mathbf{a}$ are crack parameters, and $\mathbf{K}^{b}$ are Stress Intensity Factors (SIF) along the crack front under a benchmark load $\mathbf{P}_{b}$, which is set to be 130MPa.

The reduced-order model consists of two parts: the PCA-based dimensionality reduction of crack-front nodal coordinates and the surrogate-based Stress Intensity Factor prediction, as described below.

\subsection{PCA-based Dimensionality Reduction of Crack-Front Nodal Coordinates}

In the simulation database, the vector $ \mathbf{C}_{equi, i} $, which contains the Cartesian coordinates of equispaced nodes at the crack front, is a high-dimensional representation of the crack front. Directly using $ \mathbf{C}_{equi, i} $ to construct the surrogate model would affect the computational burden and the prediction accuracy of the surrogate model. In this study, the Principal Component Analysis (PCA) is firstly conducted for the dimensionality reduction of nodal coordinates at the crack front.

In the matrix $\mathbf{C}_{equi}$, each column represents a sample crack, the total number of samples is denoted as $N$, and the number of rows $M=2p$ in which $p$ represents the number nodes along the crack front or original features. The principal component analysis is conducted to reduce the dimension of the database, in the following process.

First, the $\mathbf{C}_{equi}$ is subtracted by the mean value of the total samples $\bar{\mathbf{C}}$.

\begin{equation}
    \label{eq_5}
    \mathbf{A}=\mathbf{C}_{equi}-\bar{\mathbf{C}}_{equi}
\end{equation}

Then the Singular Value Decomposition (SVD) is executed for $\mathbf{A}$.

\begin{equation}
    \label{eq_6}
    \mathbf{A}_{m \times n}=\mathbf{U}_{M \times M} \mathbf{\Sigma}_{M \times N} \mathbf{V}^{T}_{N \times N}
\end{equation}
where $\mathbf{U} \in \mathbb{R}^{M \times M}$ and $\mathbf{V} \in \mathbb{R}^{N \times N} $ are orthogonal matrices, and $ \mathbf{\Sigma} = \operatorname{diag}_{M \times N}\left\{\sigma_{1}, \sigma_{2}, \ldots, \sigma_{N}\right\}$ contains the singular values $\sigma_{1} \geq \sigma_{2} \geq \cdots \geq \sigma_{M} \geq 0$.

After SVD of $\mathbf{A}$, the first q singular values can be used to approximate the matrix $\mathbf{A}$.
\begin{equation}
    \mathbf{A} \approx \mathbf{U}_{M \times q} \mathbf{\Sigma}_{q \times q} \mathbf{V}^{T}_{q \times N}
\end{equation}

\begin{figure}[htbp]
    \centering
    \includegraphics[width=.7\textwidth]{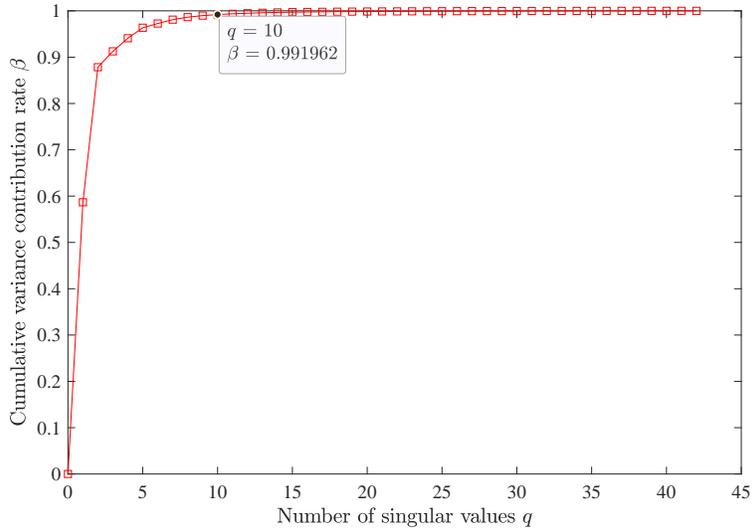}
    \caption{The relationship between the cumulative variance contribution rate $\beta$ and the number of singular values $q$ for Stage 1. When the number of singular values $q=10$, the cumulative variance contribution rate $\beta$ exceeds 0.99.}
    \label{fig: single_value}
\end{figure}

The value of $q$ is determined so that the cumulative variance contribution rate of principal components $\beta$, or the retained cumulative information, should exceed a threshold $ r $. In this paper, $ r $ is set to be 0.99. Fig. \ref{fig: single_value} shows the cumulative variance contribution rate as the number of singular values increases. When the number of singular values $q=10$, the cumulative variance contribution rate exceeds 99\%.

\begin{equation}
    \beta(q)=\frac{{\sum\limits_{{\rm{i}} = 1}^q {{\sigma _i}^2} }}{{\sum\limits_{{\rm{j}} = 1}^N {{\sigma _j}^2} }} >= r
\end{equation}

With this approximation, a crack front $\mathbf{C}_{i}$ can be projected into a $q$-dimensional subspace by
\begin{equation}
    \label{eq:projection}
    \mathbf{W}_{i} = \mathbf{U}^T_{M \times q} (\mathbf{C}_{i}-\bar{\mathbf{C}})
\end{equation}
where $\mathbf{W}_{i}$ is the principal component coefficients corresponding to the first $q$ singular values, which can also be regarded as a $q$-dimensional latent space representation of the crack front nodal coordinates $\mathbf{C}_{equi, i} $.

Conversely, the formula below can be used to restore the $\mathbf{W}_{i}$ to the full-order space.

\begin{equation}
    \label{eq_9}
    \mathbf{C}_{i} = \mathbf{U}_{M \times q}\mathbf{W}_{i}+\bar{\mathbf{C}}
\end{equation}

For all samples in the database, the principal component coefficients of each sample can be combined to form a matrix $\mathbf{W}$, which is a low-order representation of $ \mathbf{C}_{equi} $ in the database $ \mathcal{D}_{equi}$. Thus, the final database can be denoted as $ \mathcal{D}=\{ (\mathbf{W}_i,K_{i}^b):i=1,2,...,N)\}$. 

\subsection{Surrogate-Based Stress Intensity Factor Prediction}

The database $ \mathcal{D}$ is used to construct the surrogate model, in which the input is the principal component coefficients $ \mathbf{W} $ and output is the corresponding SIF. In this paper, four widely used methods are adopted to construct the models, which are least square support vector regression (LS-SVR) \cite{suykensLSSVMlabToolbox2013,suykensLeastSquaresSupport2002a}, Gaussian process regression (GPR) \cite{rasmussenGaussianProcessesMachine2010, rasmussenGaussianProcessesMachine2006}, multi-variable polynomial regression (MPR) \cite{cecenMultiPolyRegressMatlabCentral2020} and artificial neural network (ANN) \cite{pidapartiNeuralNetworkApproach1995}, respectively. For succinctness the four methods are not discussed in detail, interested readers can refer to \cite{suykensLSSVMlabToolbox2013, suykensLeastSquaresSupport2002, rasmussenGpml2018, rasmussenGaussianProcessesMachine2006, cecenAhmetcecenMultiPolyRegressMATLABCentral2020, pidapartiNeuralNetworkApproach1995} for more information.

Here, the accuracy of the four surrogate models trained with the database of the first stage (273 samples) are shown in Table \ref{tab:train and test error}. The training set consists of 205 samples and the test set consists of 68 samples. The error is defined as

\begin{equation}
    \label{eq_error}
    \varepsilon=\frac{1}{p \times N} \sum_{i=1}^{p} \sum_{j=1}^{N} \frac{\left|\hat{y}_{i j}-y_{i j}\right|}{y_{i j}}
\end{equation}
where $N$ is the number of samples in the training or test set. $ p $ is the number of crack front nodes, $\hat{y}$ is the prediction by the surrogate model and $y$ is the SIFs obtained by the original full-order simulations.

\begin{figure}[htbp]
    \centering
    \includegraphics[width=.9\textwidth]{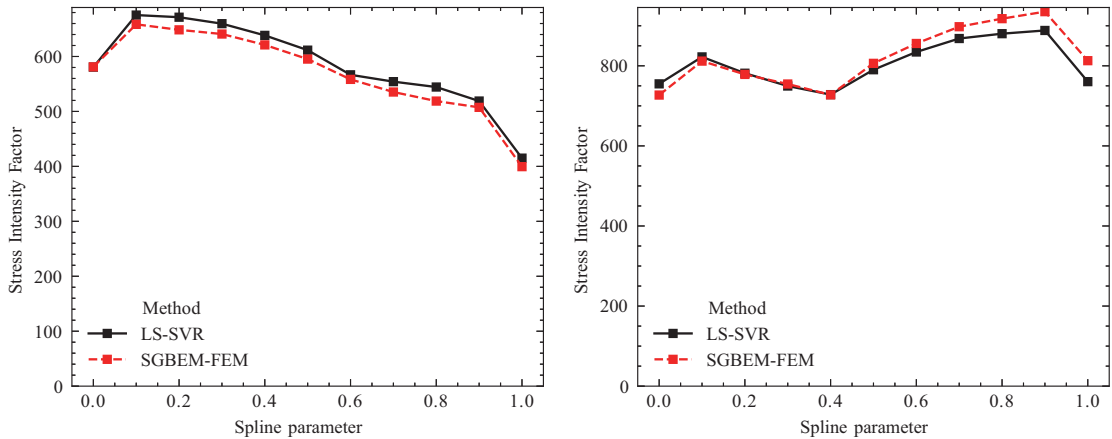}
    \caption{Prediction vs. simulation results with typical 3\% error}
    \label{fig: prediction result}
\end{figure}

The training error and test error of the four reduced-order models are listed in the Table \ref{tab:train and test error}. As can be seen, in this stage, the accuracy of the reduced-order models are similar, which are all about 3\%-5\%. Results between the LS-SVR model prediction and the original SIFs calculated by the SGBEM super element -FEM coupling method of two samples are shown in Fig. \ref{fig: prediction result}, where very good agreement is demonstrated.

\begin{table}[hbt!]
    \centering
    \caption{\label{tab:table1} Comparison of training and test error of four reduced-order models.}
    \begin{tabular}{lrrrr}
        \hline
        \rm{Model}  & \multicolumn{1}{l}{LS-SVR} & \multicolumn{1}{l}{GPR} & \multicolumn{1}{l}{MPR} & \multicolumn{1}{l}{ANN} \\
        \hline
        Training error & 0.0277                  & 0.0222                  & 0.0287                  & 0.0324                 \\
        Test error  & 0.0307                  & 0.0287                  & 0.0434                  & 0.0508                 \\
        \hline
    \end{tabular}%
    \label{tab:train and test error}%
\end{table}%

\begin{figure}[htbp]
    \centering
    \includegraphics[width=.7\textwidth]{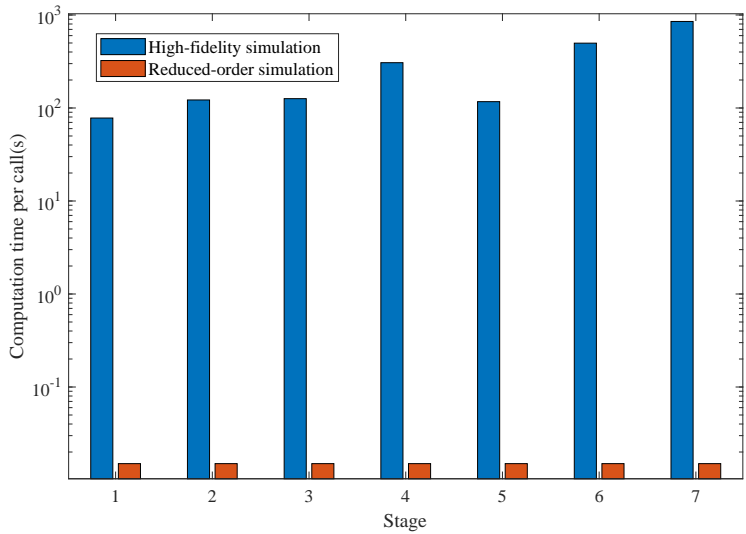}
    \caption{ Comparison of the computation time per call between the high-fidelity simulation and ROM method. The efficiency is improved by about four orders of magnitudes.}
    \label{fig: call_cmp}
\end{figure}

In addition, as a surrogate to the high-fidelity model, the computational efficiency of the reduced-order model is much higher. The typical computation time of each call in the seven different stages of the high-fidelity simulation by the SGBEM-BEM method was compared with that of the reduced order simulation shown in Fig. \ref{fig: call_cmp}. It is important to note that the computational time is plot using a logarithmic scale. The computation time per call of the reduced order model is about 0.015 seconds, which is roughly 4 orders of magnitudes smaller than the time needed for the high-fidelity Simulation.

\section{Real-time Reduced-order Crack Growth Prediction and Remaining Life Assessment of a Helicopter Component}

\label{section: application}

In this section, we firstly demonstrate how to embed the reduced order model into a deterministic crack growth prediction algorithm, which can take online monitored load data and give real-time prediction of crack growth. And then we combine this approach with Monte Carlo simulations to evaluate the probabilistic remaining life of a helicopter component, taking into account various uncertainties in crack growth.

\subsection{Deterministic Crack Growth Prediction }

Since the four regression models show comparable performance in the prediction of crack-front SIFs, the reduced-order model trained by the LS-SVR method is chosen for demonstration. In order to have an integrated algorithm of crack growth simulation, the fatigue crack growth law and boundary corrections are also needed, which are illustrated in the next two sections.

\subsubsection{Fatigue Crack Growth Law}

Fatigue crack growth laws, which relate the crack growth rate $ da/dN$ to fracture mechanics parameters (such as the range of Stress Intensity Factors $\Delta K$), can significantly influence the crack growth predictions. Widely used crack growth models include the Paris formula \cite{parisRationalAnalyticTheory1961}, the Forman formula \cite{formanNumericalAnalysisCrack1967}, the NASGRO model \cite{mettuNASGROSoftwareAnalyzing1999}, are more likely to be an empirical fitting to the experiment data. One may also empirically postulate a crack growth model from crack-growth test-data, using various data-fitting or regression methods, such as by the Moving Least Squares in \cite{dongImprovingCelebratedParis2015}. In this paper, as adopted in \cite{tiongDamageToleranceAnalysis2009}, a generalized version of the Frost–Dugdale law is used as the crack growth model:

\begin{equation}
    \label{eq: gfd}
    \frac{da}{dN}=C a^{1-\gamma/2} {\Delta \kappa}^{\gamma}
\end{equation}
where C and $\gamma$ are material constants, $ C=1.28 \times 10^{-11} $, $\gamma=3$ \cite{tiongDamageToleranceAnalysis2009} for the material used for this round-robin helicopter component. $\Delta kappa$ as the crack driving force, for which several common forms are:

\begin{equation}
    \Delta \kappa=\Delta K ;=\Delta K_{\text {eff }} ;=\Delta K^{(1-p)} K_{\max }^{p}
\end{equation}

However, as discussed in \cite{tiongDamageToleranceAnalysis2009}, since the Asterix load spectrum has only a small number of load cycles with R ratio less than 0.7, the crack driving force can be simplified as $\Delta \kappa=\Delta K$ without influencing the prediction accuracy. This simplification is adopted in this study.

\subsubsection{Boundary Correction} \label{ss:bc}

\begin{figure}[htbp]
    \centering
    \includegraphics[width=.9\textwidth]{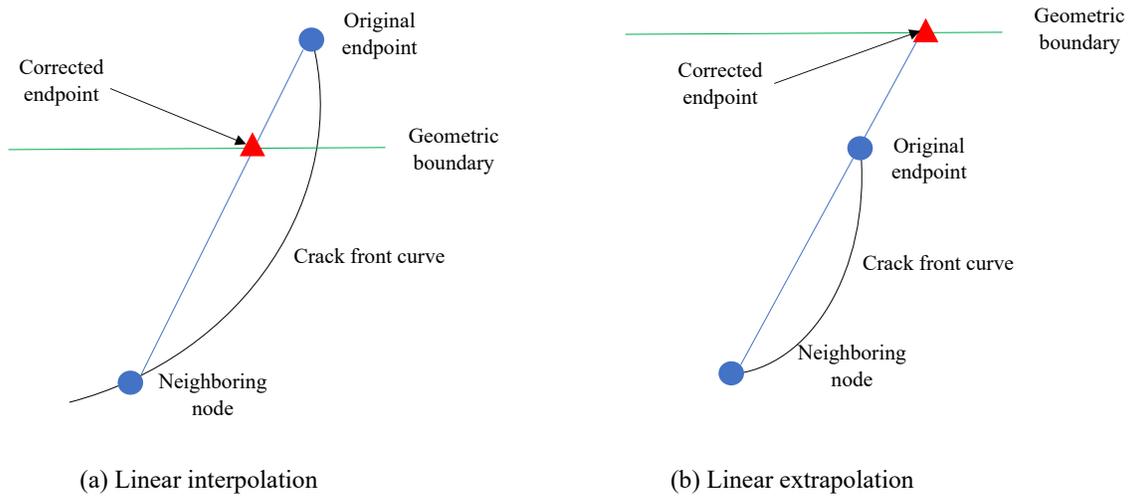}
    \caption{Schematic diagram of boundary correction. The scale is larger than it actually is for illustrating the mechanism. }
    \label{fig: boundary correction}
\end{figure}

When simulating the crack growth, the entire process is divided into several load steps. In each step, after calculating the SIFs based on the cycles of loads, the direction and size of crack growth at each node along the crack-front is calculated by the fatigue growth law, and the locations of crack-front nodes after growth is thus determined. However, the boundary of the newly determined crack front may not fall exactly on the boundary of the structural geometry. In such cases, linear interpolation and extrapolation are used to correct the boundary of the after-growth crack front curve. The schematic diagram is shown in Fig. \ref{fig: boundary correction}

\subsubsection{ The Simulation Process and Results for the Helicopter Component}

\label{section: script_simulation}
\begin{figure}[htbp]
    \centering
    \includegraphics[width=.8\textwidth]{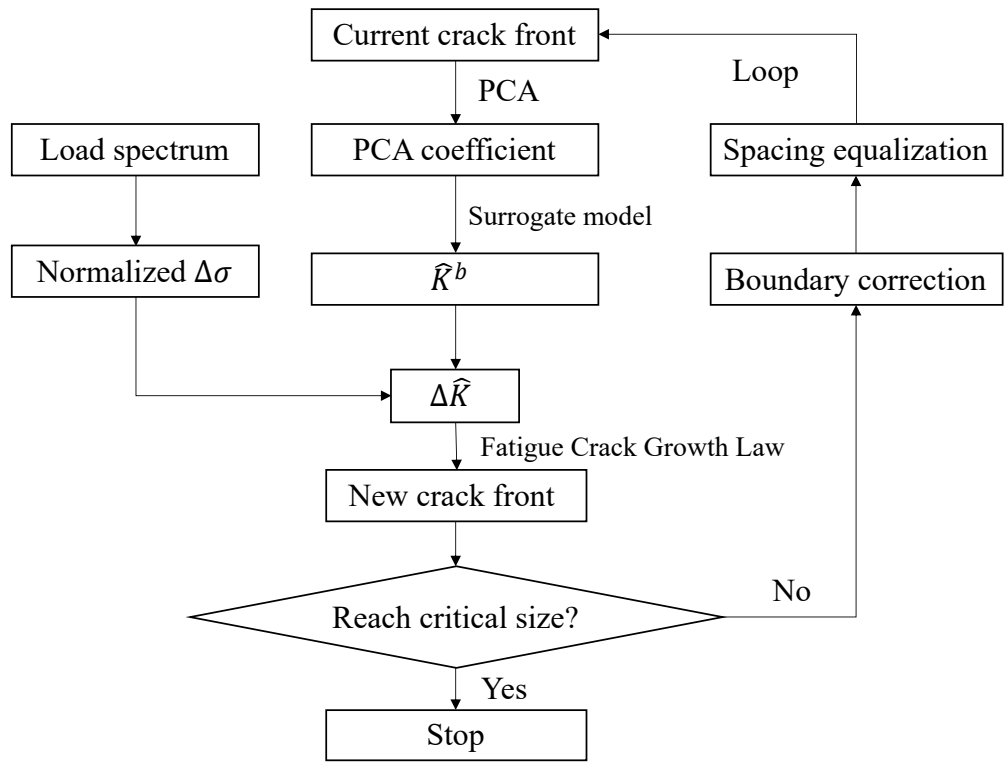}
    \caption{Flow chart of the crack growth prediction.}
    \label{fig: prediction flow}
\end{figure}

The simulation process of fatigue crack growth is illustrated in Fig. \ref{fig: prediction flow}.

The crack growth is calculated for a number of load cycles in each step. Within the step, for a crack front with equi-spaced nodes, it is first determined which stage it belongs to, and then projected to the PCA subspace by Eq. \ref{eq:projection} first to obtain principal component coefficients, which are then fed into the reduced-order fracture mechanics model to calculate the Stress Intensity Factors under the reference stress $ \hat{K}^{b}$. Then, according to the load cycle in the current step, the normalized stress range $\Delta \sigma $ can be extracted, and the SIF range $\Delta K $ can be obtained by Eq. \ref{eq: DeltaK}. It should be noted that, at stage 3 and stage 5, the lower end of the crack front is located in the lower line segment, and the two cracks may be separated. Therefore, interpolation between the two leading edges is required.

\begin{equation}
    \label{eq: DeltaK}
    \Delta \hat{K}=\Delta \sigma \cdot {\hat{K}^{b}}
\end{equation}

The crack growth increment can be calculated by the crack growth model shown in Eq. \ref{eq: gfd}. By adding the increment to the current crack front, a new crack front can be obtained. The boundary of the newly computed crack-front may not fall on the boundary of the structure, and the new nodal spacing may not be equal. The boundary correction is conducted for the new crack-front and then node equi-spacing is executed. Such a process is repeated for each load step, unless the critical crack size is reached. In this study, the critical crack size is set to be 20 mm.

Additionally, in order to increase the flexibility of online deployment, a system model of fatigue crack growth prediction was established based on Simulink. Compared with the script program, Simulink has the advantage of modularization and visualization, in which the flow of simulation data can be seen more intuitively.

In the Simulink implementation, the input is the sequence of applied loads, which is currently from the MATLAB workspace and may be instantly obtained from the data exchange interface for an online deployment in the future. The discrete state-space block is used to represent the development of the crack front. The dimension of the state variables is 43, where the first 42 variables are the coordinates of the 21 crack front nodes, and the last value is used to indicate whether the crack front is split. A MATLAB function is written for the selection of the crack growth stage, which is based on the coordinates of the upper and lower edge nodes, followed by the PCA projector to projects the crack front coordinates into the PCA coefficients. Considering the different stages and whether the front is split, a total of nine feed-forward neural network models, exported from the MATLAB artificial neural network toolbox, are implemented, where the switching is realized through the multiport switch block. When the Stress Intensity Factors under the reference load is obtained using the surrogate model, the corresponding stress intensity factors in the current load cycle can be obtained by the product block implemented Eq. \ref{eq: DeltaK}. Based on the MATLAB function block, the crack growth module and boundary correction module are implemented respectively. In order to control the ending time of the simulation, a switch block and a stop block is deployed, by which the simulation is stopped when the surface crack reaches 20 mm. The complete process is shown in Fig. 15, and the information of various modules is shown in Table \ref{tab:system_modules}.

\begin{figure}[htbp]
    \centering
    \includegraphics[width=1.0\textwidth]{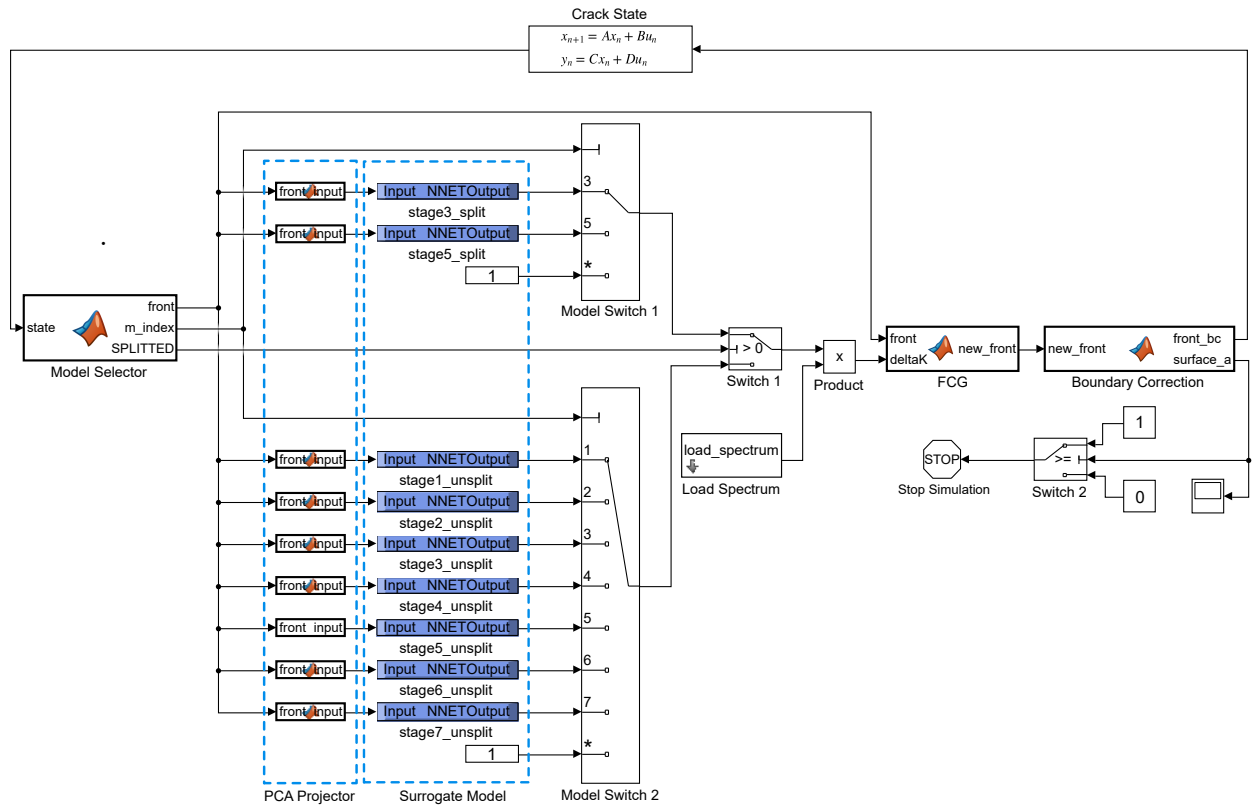}
    \caption{Schematic diagram of the system simulation of the fatigue crack growth}
    \label{fig: system_simulation}
\end{figure}

\begin{table}[htbp]
    \centering
    \caption{Main modules of the system model of the fatigue crack growth}
      \begin{tabular}{lp{3.3cm}cp{7cm}}
        \hline
        Module name & Type  & Count & Description \\
        \hline
        Crack State & Discrete State-space & 1     & Store the state of the crack front \\
        Model Selector & MATLAB Function & 1     & Determine the crack stage to select an appropriate ROM \\
        Switch 1 & Switch & 1 & Switch to corresponding ROMs set by the indicator of SPLITTED \\
        Model Switch & Multiport Switch & 2     & Switch to corresponding ROM according to m\_index\\
        PCA Projector & MATLAB Function & 9     & Project the crack front into a low-dimensional subspace as the input of the surrogate model \\
        Surrogate Model & Feed-Forward Neural Network & 9 & Predict the Stress Intensity Factors under the reference stress $ \hat{K}^{b}$  \\
        Load Spectrum & Signal From Workspace & 1     & Provide the stress range $\Delta \sigma $ of each load cycle \\
        Product & Product & 1     & The $ \hat{K}^{b}$ and $\Delta \sigma $ are multiplied to obtain SIF range $ \Delta \hat{K} $ \\
        FCG & MATLAB Function & 1     & Predict the new crack front by the fatigue crack growth law \\
        Boundary Correction & MATLAB Function & 1 & Correct the boundary of the new  crack front\\        
        Switch 2 & Switch & 1 & Indicate whether a critical crack length has been reached \\
        Stop Simulation & Stop & 1 & Stop the fatigue crack growth simulation \\
      \hline
      \end{tabular}%
    \label{tab:system_modules}%
\end{table}%

\begin{figure}[htbp]
    \centering
    \includegraphics[width=.95\textwidth]{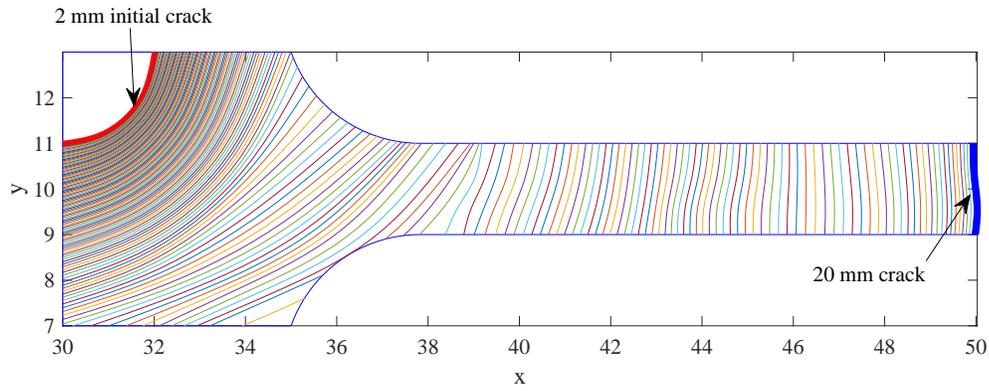}
    \caption{Crack growth path from 2 mm to 20 mm. The initial crack is a 2 mm corner crack at the upper left corner of the figure (red curve), and the  20 mm crack is in the right side of the figure (blue line).}
    \label{fig: crack_growth_path}
\end{figure}

The predicted path of crack growths from 2 mm to 20 mm by the reduced-order model is shown in Fig. \ref{fig: crack_growth_path}. It can be seen that, starting from a 2 mm corner crack, the crack gradually grows across the thickness and then transforms into a through-thickness crack. 
 
\begin{figure}[htbp]
    \centering
    \includegraphics[width=.8\textwidth]{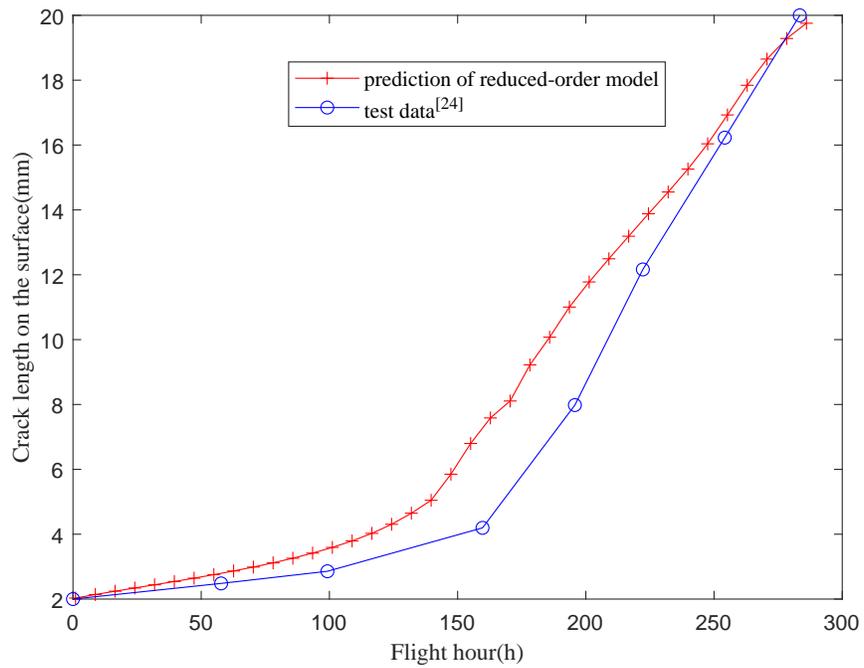}
    \caption{Comparison of the reduced-order model predictions and round-robin test data\cite{cansdaleHelicopterDamageTolerance2002}}
    \label{fig: crack_growth_time}
\end{figure}

The comparison of the crack growth simulation and the round-robin test data is shown in Fig. \ref{fig: crack_growth_time}, from which we can see that fairly good agreements are obtained. In the first 150 hours, the predicted growth is slightly faster than that in the test, while after that, the predicted crack growth is slightly slower. 

\subsubsection{Influence of the load step size}

In terms of the computation time, when the step size is set to 1, which means the calculation is conducted cycle by cycle of about half a million cycles, the computation time is only 20 minutes on an ordinary desktop PC. In this case, each call of the reduced-order model takes about 0.015 seconds, which is about four orders of magnitude less than that of the original SGBEM-FEM simulation. If the step size is set to be 1000, the entire simulation takes only about 12 seconds. In order to accelerate the simulation, using a larger load step size is beneficial. 

\begin{figure}[htbp]
    \centering
    \subfigure[crack growth simulation considering different load steps.]{
        \includegraphics[width=.50\textwidth]{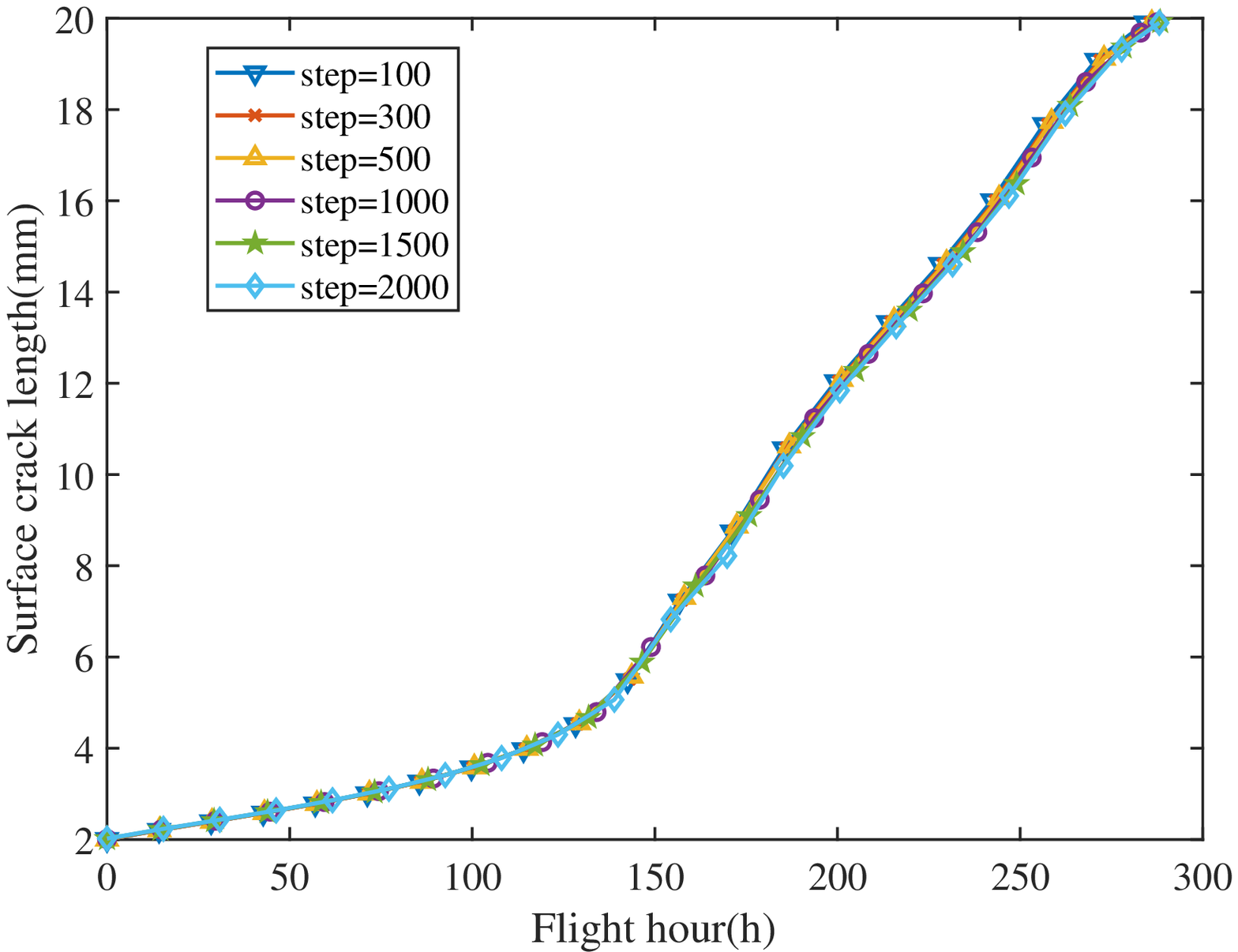}}
    \subfigure[The simulated crack growth life considering different load steps]{
        \includegraphics[width=.47\textwidth]{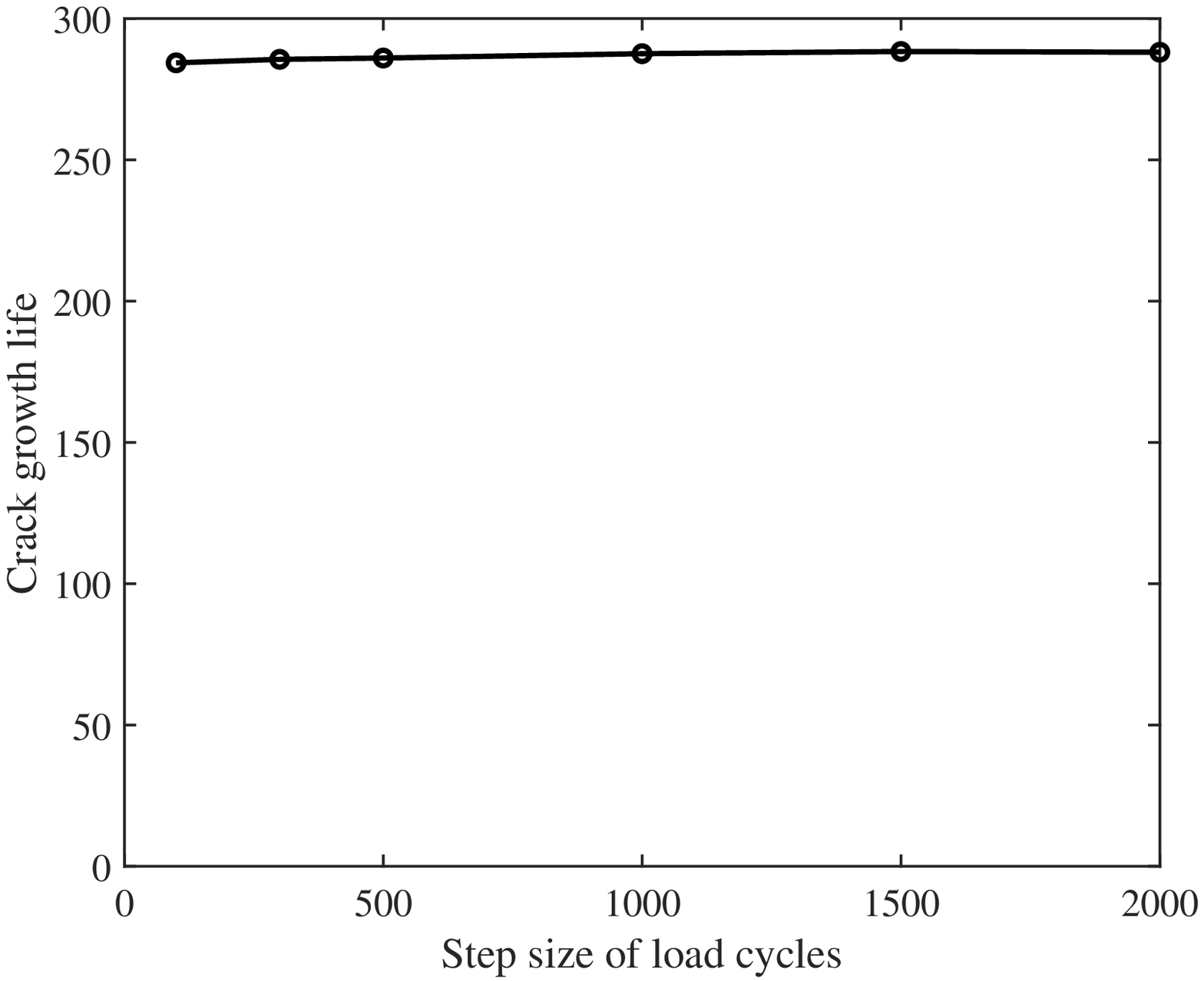}}%
    \caption{Comparison of the crack growth and remaining life under different load steps}
    \label{fig: crack_growth_under_diff_steps}
\end{figure}

In order to study the effect of the step size on the simulation accuracy, six step sizes are adopted for the crack growth simulation which are 100,300,500,1000,1500, 2000, respectively. The result is shown in Fig. \ref{fig: crack_growth_under_diff_steps}. It can be seen that the influence of the step size is very small. Keeping in mind that with a load step of 1000 cycles, the entire simulation is completed in about 12 seconds, we may say that the simulation by this approach is faster than real-time (as compared to the physical test which takes slightly less than 300 hours).

\subsection{ Probabilistic Assessment of the Crack-growth Life Considering Uncertainties}
\label{section: stochastic simulations}

For the deployment of a structural digital twin, prior distributions of various uncertainties, such as parameters of the fatigue crack growth model, and size/ shape of the initial crack, need to be specified to conduct probabilistic crack growth simulations. These uncertainties will significantly affect the predicted remaining useful life of the structural component, and they should be continuously tracked and controlled by the fusion of probabilistic predictions and monitoring/ inspection data in the following service stage. The reduced-order modeling method proposed in this paper provides a powerful tool to reduce the computational burden of probabilistic crack growth simulations. In this subsection, the following uncertain parameters are considered for demonstration:
\begin{enumerate}
    \item parameters of the generalized Frost-Dugdale law: C and $\gamma$.
    \item shape parameters of the initial corner crack $L_{a}$ and $L_{c}$.
\end{enumerate}

\begin{table}[htbp]
    \centering
    \caption{Prior distribution of uncertain parameters}
      \begin{tabular}{llr}
        \hline
        Parameters & Distribution & \multicolumn{1}{l}{Description} \\
        \hline
        C     & U(0.28E-11,1.58E-11) & Uniform distribution\\
        $\gamma$ & N(3,0.01) &  Normal distribution\\
        $L_a$  & U(1.5,2.5) &  Uniform distribution\\
        $L_c$  & U(1.5,2.5) &  Uniform distribution\\
        \hline
      \end{tabular}
    \label{tab:prior}
\end{table}

\begin{figure}[htbp]
    \centering
    \subfigure[Crack growth predictions.]{
        \includegraphics[width=.48\textwidth]{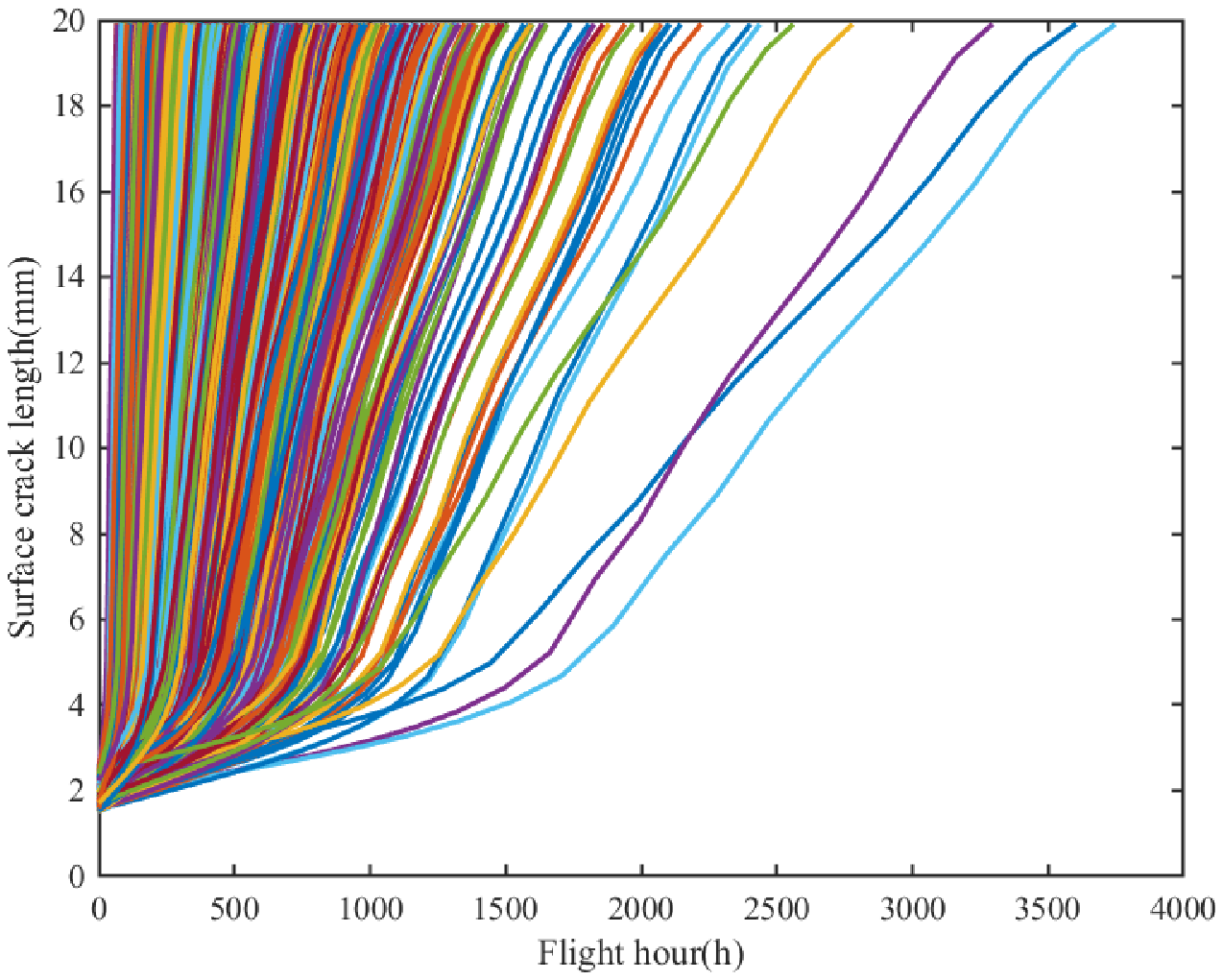}}
    \subfigure[Distribution of the computed life, the mean value is 422.9h, and the median value is 309.5h.]{
        \includegraphics[width=.48\textwidth]{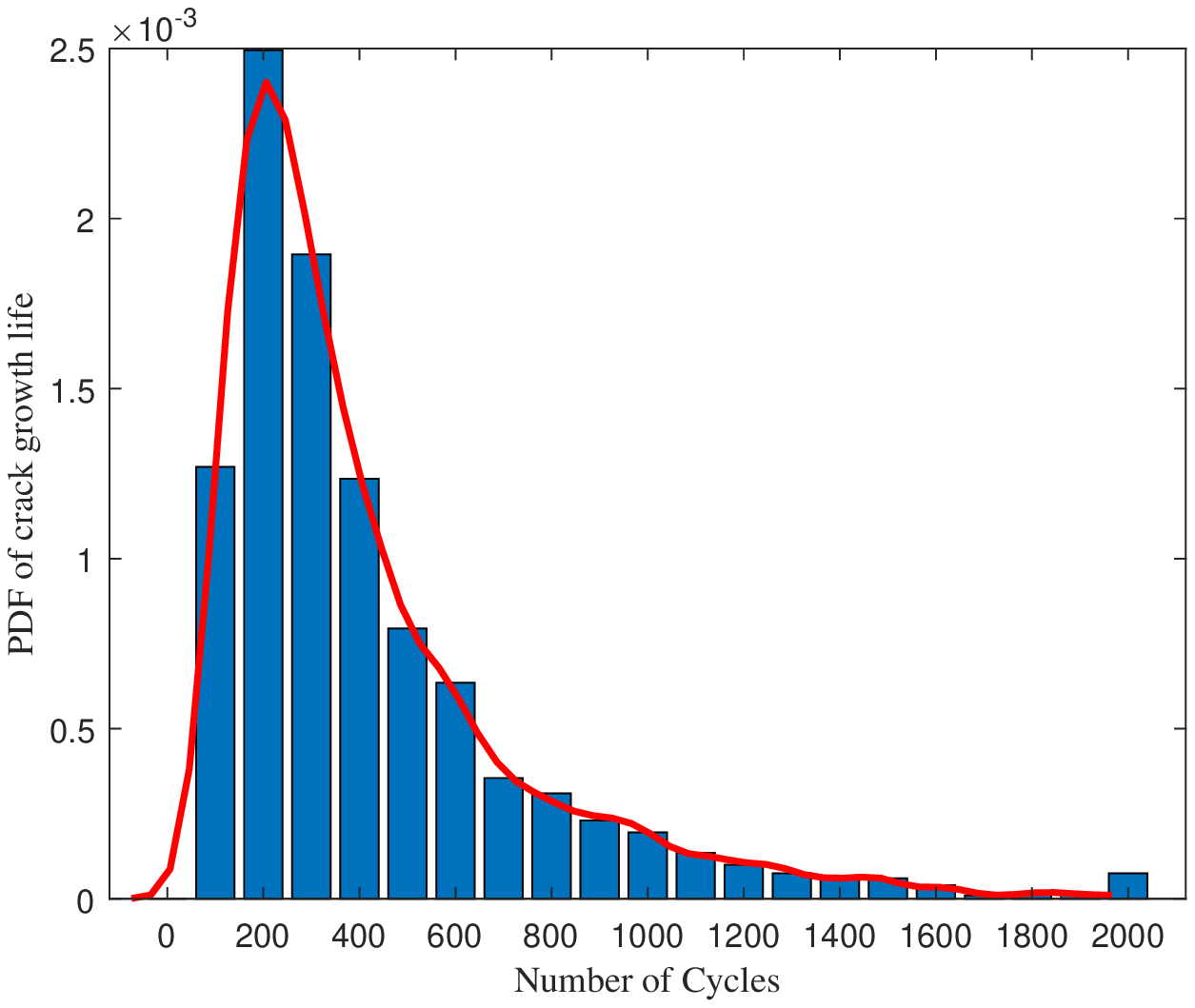}}%
    \caption{Crack growth curve and remaining life distribution by the Monte Carlo simulation}
    \label{fig: simulation_2000_samples}
\end{figure}

\begin{figure}[htbp]
    \centering
    \includegraphics[width=.8\textwidth]{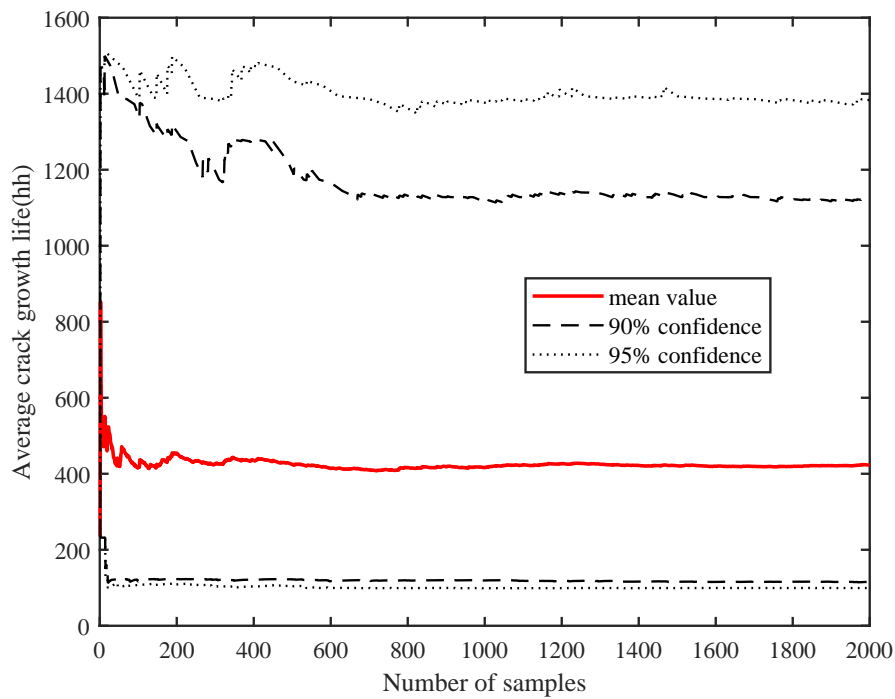}
    \caption{Effect of the number of samples on the crack growth life distribution}
    \label{fig: effect_samples}
\end{figure}

A Monte Carlo simulation is conducted to demonstrate the capacity of the reduced order modeling approach in the probabilistic remaining life assessment. 2000 samples with different fatigue growth law parameters and different initial crack sizes are generated by the Latin hypercube sampling from the prior distribution shown in Table \ref{tab:prior}, and crack-growth predictions are carried out for each sample. The crack growth predictions and the computed distribution of the computed life distribution are shown in Fig \ref{fig: simulation_2000_samples}. The computed life of different samples varies from 64.5714h to 3477.8h, which demonstrates the influence of uncertain parameters. It can be seen that the distribution of crack-growth life has a long tail effect, which is caused by a small number of samples with relatively small $\gamma$ and thus have a very large crack growth life. Regarding the effect of the number of samples on the computed distribution of the crack-growth life, as can be seen from Fig. \ref{fig: effect_samples}, when the number of samples exceeds 800, the computed mean value, 90\% and 95\% confidence bounds are overall very stable.

It should be noted that, with the reduced-order crack growth simulation tool proposed in this paper, the fusion of probabilistic crack-growth prediction and multi-source data obtained from an in-service structural component can be done quite easily. Under the framework of digital twin, utilizing traditional visual inspection data, nondestructive testing data, structural health monitoring (SHM) data\cite{sbarufattiApplicationSensorTechnologies2014, wangUseHighPerformanceFatigue2015} as well as potential early fatigue signatures/ damage precursors precursor\cite{henryEvaluationEarlyFatigue2020} may effectively track and control the uncertainties for the fatigue crack growth\cite{wangUseHighPerformanceFatigue2015}. This will be studied in our future work.

\section{Conclusion}
\label{section: conclusion}

This paper proposes a real-time simulation approach to conduct deterministic/probabilistic predictions of crack growth and the remaining life of a structural component. This approach is featured by combining the advantages of the SGBEM super element - FEM coupling method for fracture mechanics simulations to construct an offline database, and advantages of the reduced-order modeling approach to accelerate the online simulation efficiently. Compared with existing methods, the proposed method is advantageous for the real-time simulation of the growth of complex-shaped cracks in complex structures, which has potential applications in constructing digital twins of complex engineering structures.

The SGEBM super element - FEM coupling method combined with a parametric modeling method are employed to generate the database of SIFs of different cracks in a complex round-robin helicopter component. It builds the FEM model of the uncracked global structure and the SGBEM model of a cracked subdomain separately, and couples them by the assembly of stiffness matrices in the simulation process. It significantly reduces the human labor for preprocessing, and improves the computational efficiency for full-order fracture mechanics simulations.
 
Principal Component Analysis and four surrogate modeling methods are used to construct reduced-order models from the database, to replace full-order fracture mechanics simulations. Using the reduced-order model, the predicted crack growth path and life of the helicopter component under a Asterix load spectrum are close to the test results. To increase the flexibility of online deployment, a system model of fatigue crack growth prediction was established based on Simulink. Further, a simple Monto-Carlo simulation is conducted to evaluate the remaining life distribution of the component. Real-time simulations can be realized by using the reduced order model (as compared to the time required for the physical test). 
 
It is expected that the methodology can be easily extended to non-planar growth of crack surfaces and crack-fronts under more complex mixed mode loadings. Furthermore, by utilizing the data of load/usage monitoring and crack inspections for a servicing structural component, the proposed reduced modeling method can be used for the deployment of helicopter component digital twin, leading to an intelligent framework of inspection and repair. This will be the focus of our future studies. 

\section*{ Funding Sources}

The work of the first 3 authors was supported by the National Natural Science Foundation of China (grant number 12072011), the Aeronautical Science Foundation of China (grant number 201909051001) and the Seed Foundation of Beijing Advanced Discipline Center for Unmanned Aircraft System (grant number ADBUAS-2019-SP-05).

\section*{ Acknowledgments}

The authors express gratitude to Rhys Jones at Monash University for providing the Asterix spectrum.

\bibliography{reference}

\begin{thebibliography}{41}
\newcommand{\enquote}[1]{``#1''}
\providecommand{\natexlab}[1]{#1}
\providecommand{\url}[1]{\texttt{#1}}
\providecommand{\urlprefix}{URL }
\expandafter\ifx\csname urlstyle\endcsname\relax
  \providecommand{\doi}[1]{\discretionary{}{}{}https://doi.org/#1}\else
  \providecommand{\doi}[1]{\discretionary{}{}{}\urlstyle{rm}\url{https://doi.org/#1}}\fi

\bibitem[{Miner(1945)}]{minerCumulativeDamageFatigue1945}
Miner, M., \enquote{Cumulative {{Damage}} in {{Fatigue}},} \emph{Journal of
  Applied Mechanics}, Vol.~12, No.~3, 1945, pp. A159--A164.
\newblock \doi{10.1115/1.4009458}.

\bibitem[{Zhao and Adams(2011)}]{zhaoChallengesDamageTolerance2011}
Zhao, J., and Adams, D., \enquote{Challenges in {{Damage Tolerance Approach}}
  for {{Dynamic Loaded Rotorcraft Components}} \textendash{} {{From Risk
  Assessment}} to {{Optimal Inspection Planning}},} \emph{{{ICAF}} 2011
  {{Structural Integrity}}: {{Influence}} of {{Efficiency}} and {{Green
  Imperatives}}}, edited by J.~Komorowski, {Springer Netherlands}, {Dordrecht,
  The Netherlands}, 2011, pp. 927--957.
\newblock \doi{10.1007/978-94-007-1664-3_71}.

\bibitem[{Grieves(12, 2014)}]{grievesDigitalTwinManufacturing2014}
Grieves, M., \enquote{Digital {{Twin}}: {{Manufacturing Excellence}} through
  {{Virtual Factory Replication}},} \emph{White paper}, Vol.~1, 12, 2014, pp.
  1--7.

\bibitem[{Cheung and Bellinger(2009)}]{cheungRoadmapHelicopterLife2009}
Cheung, C., and Bellinger, N.~C., \enquote{Roadmap for {{Helicopter Life Usage
  Monitoring}},} \emph{Thirteenth {{Australian International Aerospace
  Congress}}}, {Royal Aeronautical Society}, {Melbourne}, 2009, pp. 1--13.

\bibitem[{Tuegel et~al.(2011)Tuegel, Ingraffea, Eason, and
  Spottswood}]{tuegelReengineeringAircraftStructural2011}
Tuegel, E.~J., Ingraffea, A.~R., Eason, T.~G., and Spottswood, S.~M.,
  \enquote{Reengineering {{Aircraft Structural Life Prediction Using}} a
  {{Digital Twin}},} \emph{International Journal of Aerospace Engineering},
  Vol. 2011, 2011, p. 154798.
\newblock \doi{10.1155/2011/154798}.

\bibitem[{Leser et~al.(2020)Leser, Warner, Leser, Bomarito, Newman, and
  Hochhalter}]{leserDigitalTwinFeasibility2020}
Leser, P.~E., Warner, J.~E., Leser, W.~P., Bomarito, G.~F., Newman, J.~A., and
  Hochhalter, J.~D., \enquote{A Digital Twin Feasibility Study ({{Part II}}):
  {{Non-deterministic}} Predictions of Fatigue Life Using in-Situ Diagnostics
  and Prognostics,} \emph{Engineering Fracture Mechanics}, Vol. 229, 2020, p.
  106903.
\newblock \doi{10.1016/j.engfracmech.2020.106903}.

\bibitem[{Renaud et~al.(2020)Renaud, Liao, and
  Bombardier}]{renaudDemonstrationAirframeDigital2020}
Renaud, G., Liao, M., and Bombardier, Y., \enquote{Demonstration of an
  {{Airframe Digital Twin Framework Using}} a {{CF-188 Full-Scale Component
  Test}},} \emph{{{ICAF}} 2019 \textendash{} {{Structural Integrity}} in the
  {{Age}} of {{Additive Manufacturing}}}, edited by A.~Niepokolczycki and
  J.~Komorowski, {Springer International Publishing}, {Cham, Switzerland},
  2020, pp. 176--186.

\bibitem[{Liao et~al.(2020)Liao, Renaud, and
  Bombardier}]{liaoAirframeDigitalTwin2020}
Liao, M., Renaud, G., and Bombardier, Y., \enquote{Airframe Digital Twin
  Technology Adaptability Assessment and Technology Demonstration,}
  \emph{Engineering Fracture Mechanics}, Vol. 225, 2020, p. 106793.
\newblock \doi{10.1016/j.engfracmech.2019.106793}.

\bibitem[{West and Blackburn(2017)}]{westDigitalThreadDigital2017}
West, T.~D., and Blackburn, M., \enquote{Is {{Digital Thread}}/{{Digital Twin
  Affordable}}? {{A Systemic Assessment}} of the {{Cost}} of {{DoD}}'s {{Latest
  Manhattan Project}},} \emph{Procedia Computer Science}, Vol. 114, 2017, pp.
  47--56.
\newblock \doi{10.1016/j.procs.2017.09.003}.

\bibitem[{Cerrone et~al.(2014)Cerrone, Hochhalter, Heber, and
  Ingraffea}]{cerroneEffectsModelingAsManufactured2014}
Cerrone, A., Hochhalter, J., Heber, G., and Ingraffea, A., \enquote{On the
  {{Effects}} of {{Modeling As-Manufactured Geometry}}: {{Toward Digital
  Twin}},} \emph{International Journal of Aerospace Engineering}, Vol. 2014,
  2014, p. 439278.
\newblock \doi{10.1155/2014/439278}.

\bibitem[{Sankararaman et~al.(2011)Sankararaman, Ling, Shantz, and
  Mahadevan}]{sankararamanUncertaintyQuantificationFatigue2011}
Sankararaman, S., Ling, Y., Shantz, C., and Mahadevan, S., \enquote{Uncertainty
  {{Quantification}} in {{Fatigue Crack Growth Prognosis}},}
  \emph{International Journal of Prognostics and Health Management}, Vol.~2,
  No.~1, 2011, pp. 1--15.

\bibitem[{Hombal and Mahadevan(2013)}]{hombalSurrogateModeling3D2013}
Hombal, V., and Mahadevan, S., \enquote{Surrogate Modeling of {{3D}} Crack
  Growth,} \emph{International Journal of Fatigue}, Vol.~47, 2013, pp. 90--99.
\newblock \doi{10.1016/j.ijfatigue.2012.07.012}.

\bibitem[{Leser et~al.(2017)Leser, Hochhalter, Warner, Newman, Leser,
  Wawrzynek, and Yuan}]{leserProbabilisticFatigueDamage2017}
Leser, P.~E., Hochhalter, J.~D., Warner, J.~E., Newman, J.~A., Leser, W.~P.,
  Wawrzynek, P.~A., and Yuan, F.-G., \enquote{Probabilistic Fatigue Damage
  Prognosis Using Surrogate Models Trained via Three-Dimensional Finite Element
  Analysis,} \emph{Structural Health Monitoring}, Vol.~16, No.~3, 2017, pp.
  291--308.
\newblock \doi{10.1177/1475921716643298}.

\bibitem[{Keprate et~al.(2017{\natexlab{a}})Keprate, Ratnayake, and
  Sankararaman}]{keprateAdaptiveGaussianProcess2017}
Keprate, A., Ratnayake, R.~C., and Sankararaman, S., \enquote{Adaptive
  {{Gaussian}} Process Regression as an Alternative to {{FEM}} for Prediction
  of Stress Intensity Factor to Assess Fatigue Degradation in Offshore
  Pipeline,} \emph{International Journal of Pressure Vessels and Piping}, Vol.
  153, 2017{\natexlab{a}}, pp. 45--58.
\newblock \doi{10.1016/j.ijpvp.2017.05.010}.

\bibitem[{Keprate et~al.(2017{\natexlab{b}})Keprate, Ratnayake, and
  Sankararaman}]{keprateSurrogateModelPredicting2017}
Keprate, A., Ratnayake, R. M.~C., and Sankararaman, S., \enquote{A {{Surrogate
  Model}} for {{Predicting Stress Intensity Factor}}: {{An Application}} to
  {{Oil}} and {{Gas Industry}},} \emph{{{ASME}} 2017 36th {{International
  Conference}} on {{Ocean}}, {{Offshore}} and {{Arctic Engineering}}}, Vol.~4,
  {American Society of Mechanical Engineers}, {Trondheim, Norway},
  2017{\natexlab{b}}, pp. 1--12.
\newblock \doi{10.1115/OMAE2017-61091}.

\bibitem[{Li et~al.(2017)Li, Mahadevan, Ling, Choze, and
  Wang}]{liDynamicBayesianNetwork2017}
Li, C., Mahadevan, S., Ling, Y., Choze, S., and Wang, L., \enquote{Dynamic
  {{Bayesian Network}} for {{Aircraft Wing Health Monitoring Digital Twin}},}
  \emph{AIAA Journal}, Vol.~55, No.~3, 2017, pp. 930--941.
\newblock \doi{10.2514/1.J055201}.

\bibitem[{Nishioka and Atluri(1983)}]{nishiokaAlternatingMethodAnalysis1983}
Nishioka, T., and Atluri, S.~N., \enquote{An Alternating Method for Analysis of
  Surface-Flawed Aircraft Structural Components,} \emph{AIAA Journal}, Vol.~21,
  No.~5, 1983, pp. 749--757.
\newblock \doi{10.2514/3.8143}.

\bibitem[{Park and Atluri(1998)}]{parkMixedModeFatigue1998}
Park, J.~H., and Atluri, S.~N., \enquote{Mixed Mode Fatigue Growth of Curved
  Cracks Emanating from Fastener Holes in Aircraft Lap Joints,}
  \emph{Computational Mechanics}, Vol.~21, No.~6, 1998, pp. 477--482.
\newblock \doi{10.1007/s004660050326}.

\bibitem[{Nikishkov et~al.(2001)Nikishkov, Park, and
  Atluri}]{nikishkovSGBEMFEMAlternatingMethod2001}
Nikishkov, G., Park, J.-H., and Atluri, S., \enquote{{{SGBEM-FEM Alternating
  Method}} for {{Analyzing 3D Non-Planar Cracks}} and {{Their Growth}} in
  {{Structural Components}},} \emph{CMES: Computer Modeling in Engineering \&
  Sciences}, Vol.~2, No.~3, 2001, pp. 401--422.

\bibitem[{Han and Atluri(2002)}]{hanSGBEMCrackedLocal2002}
Han, Z.~D., and Atluri, S.~N., \enquote{{{SGBEM}} (for {{Cracked Local
  Subdomain}}) \textendash{} {{FEM}} (for Uncracked Global {{Structure}})
  {{Alternating Method}} for {{Analyzing 3D Surface Cracks}} and {{Their
  Fatigue-Growth}},} \emph{CMES: Computer Modeling in Engineering \& Sciences},
  Vol.~3, No.~6, 2002, pp. 699--716.

\bibitem[{Dong and Atluri(2012)}]{dongSGBEMUsingNonhypersingular2012}
Dong, L., and Atluri, S.~N., \enquote{{{SGBEM}} ({{Using Non-hyper-singular
  Traction BIE}}), and {{Super Elements}}, for {{Non-Collinear Fatigue-growth
  Analyses}} of {{Cracks}} in {{Stiffened Panels}} with {{Composite-Patch
  Repairs}},} \emph{CMES: Computer Modeling in Engineering \& Sciences},
  Vol.~89, No.~5, 2012, pp. 415--456.

\bibitem[{Dong and
  Atluri(2013{\natexlab{a}})}]{dongFractureFatigueAnalyses2013a}
Dong, L., and Atluri, S.~N., \enquote{Fracture \& Fatigue Analyses:
  {{SGBEM-FEM}} or {{XFEM}}? {{Part}} 1: {{2D}} Structures,} \emph{CMES:
  Computer Modeling in Engineering \& Sciences}, Vol.~90, No.~2,
  2013{\natexlab{a}}, pp. 91--146.

\bibitem[{Dong and
  Atluri(2013{\natexlab{b}})}]{dongFractureFatigueAnalyses2013}
Dong, L., and Atluri, S.~N., \enquote{Fracture \& Fatigue Analyses:
  {{SGBEM-FEM}} or {{XFEM}}? {{Part}} 2: {{3D}} Solids,} \emph{CMES: Computer
  Modeling in Engineering \& Sciences}, Vol.~90, No.~5, 2013{\natexlab{b}}, pp.
  379--413.

\bibitem[{Cansdale and Perrett(2002)}]{cansdaleHelicopterDamageTolerance2002}
Cansdale, R., and Perrett, B., \enquote{The Helicopter Damage Tolerance Round
  Robin Challenge,} \emph{Workshop on Fatigue Damage of Helicopters}, Vol.~12,
  {University of Pisa}, {Pisa, Italy}, 2002, pp. 99--128.

\bibitem[{Irving et~al.(2003)Irving, Lin, and
  Bristow}]{irvingDamageToleranceHelicopters2003}
Irving, P., Lin, J., and Bristow, J., \enquote{Damage {{Tolerance}} in
  {{Helicopters Report}} on the {{Round Robin Challenge}},} \emph{59th
  {{American Helicopter Society Annual Forum}}}, {Vertical Flight Society},
  {Phoenix, Arizona}, 2003, pp. 1642--1652.

\bibitem[{Newman et~al.(2006)Newman, Irving, Lin, and
  Le}]{newmanCrackGrowthPredictions2006}
Newman, J.~C., Irving, P.~E., Lin, J., and Le, D.~D., \enquote{Crack Growth
  Predictions in a Complex Helicopter Component under Spectrum Loading,}
  \emph{Fatigue Fracture of Engineering Materials and Structures}, Vol.~29,
  No.~11, 2006, pp. 949--958.
\newblock \doi{10.1111/j.1460-2695.2006.01053.x}.

\bibitem[{Tiong and Jones(2009)}]{tiongDamageToleranceAnalysis2009}
Tiong, U., and Jones, R., \enquote{Damage Tolerance Analysis of a Helicopter
  Component,} \emph{International Journal of Fatigue}, Vol.~31, No.~6, 2009,
  pp. 1046--1053.
\newblock \doi{10.1016/j.ijfatigue.2008.05.012}.

\bibitem[{Vaughan and Chang(2003)}]{vaughanLifePredictionHigh2003}
Vaughan, R.~E., and Chang, J.~H., \enquote{Life Prediction for High Cycle
  Dynamic Components Using Damage Tolerance and Small Threshold Cracks,}
  \emph{59th {{American Helicopter Society Annual Forum}}}, Vol.~3, {Vertical
  Flight Society}, {Phoenix, Arizona}, 2003, pp. 1712--1720.

\bibitem[{Suykens(2013)}]{suykensLSSVMlabToolbox2013}
Suykens, J., \enquote{{{LS-SVMlab}} Toolbox,} ESAT-SISTA, Sep. 2013.

\bibitem[{Suykens et~al.(2002)Suykens, Van~Gestel, De~Brabanter, De~Moor, and
  Vandewalle}]{suykensLeastSquaresSupport2002a}
Suykens, J.~A., Van~Gestel, T., De~Brabanter, J., De~Moor, B., and Vandewalle,
  J.~P., \emph{Least Squares Support Vector Machines}, {World scientific},
  {River Edge, NJ}, 2002.

\bibitem[{Rasmussen and
  Nickisch(11,2010)}]{rasmussenGaussianProcessesMachine2010}
Rasmussen, C.~E., and Nickisch, H., \enquote{Gaussian Processes for Machine
  Learning ({{GPML}}) Toolbox,} \emph{The Journal of Machine Learning
  Research}, Vol.~11, 11,2010, pp. 3011--3015.

\bibitem[{Rasmussen and Williams(2006)}]{rasmussenGaussianProcessesMachine2006}
Rasmussen, C.~E., and Williams, C. K.~I., \emph{Gaussian Processes for Machine
  Learning}, Adaptive Computation and Machine Learning, {MIT Press},
  {Cambridge, Mass}, 2006.

\bibitem[{Cecen(2020)}]{cecenMultiPolyRegressMatlabCentral2020}
Cecen, A., \enquote{{{MultiPolyRegress-MatlabCentral}},} MINED@Gatech, Sep.
  2020.

\bibitem[{Pidaparti and Palakal(1995)}]{pidapartiNeuralNetworkApproach1995}
Pidaparti, R. M.~V., and Palakal, M.~J., \enquote{Neural Network Approach to
  Fatigue-Crack-Growth Predictions under Aircraft Spectrum Loadings,}
  \emph{Journal of Aircraft}, Vol.~32, No.~4, 1995, pp. 825--831.
\newblock \doi{10.2514/3.46797}.

\bibitem[{Paris et~al.(1961)Paris, Gomez, and
  Anderson}]{parisRationalAnalyticTheory1961}
Paris, P.~C., Gomez, M.~P., and Anderson, W. E.~P., \enquote{A {{Rational
  Analytic Theory}} of {{Fatigue}},} \emph{Trend in Engineering}, Vol.~13,
  1961, pp. 9--14.

\bibitem[{Forman et~al.(1967)Forman, Kearney, and
  Engle}]{formanNumericalAnalysisCrack1967}
Forman, R.~G., Kearney, V.~E., and Engle, R.~M., \enquote{Numerical
  {{Analysis}} of {{Crack Propagation}} in {{Cyclic-Loaded Structures}},}
  \emph{Journal of Basic Engineering}, Vol.~89, No.~3, 1967, pp. 459--463.
\newblock \doi{10.1115/1.3609637}.

\bibitem[{Mettu et~al.(1999)Mettu, Shivakumar, Beek, Yeh, Williams, Forman,
  McMahon, and Newman~Jr}]{mettuNASGROSoftwareAnalyzing1999}
Mettu, S.~R., Shivakumar, V., Beek, J.~M., Yeh, F., Williams, L.~C., Forman,
  R.~G., McMahon, J.~J., and Newman~Jr, J.~C., \enquote{{{NASGRO}} 3.0: {{A}}
  Software for Analyzing Aging Aircraft,} \emph{The {{Second Joint
  NASA}}/{{FAA}}/{{DoD Conference}} on {{Aging Aircraft}}}, {National
  Aeronautics and Space Administration}, {Williamsburg, VA}, 1999, pp.
  792--801.

\bibitem[{Dong et~al.(2015)Dong, Haynes, and
  Atluri}]{dongImprovingCelebratedParis2015}
Dong, L., Haynes, R., and Atluri, S.~N., \enquote{On {{Improving}} the
  {{Celebrated Paris}}' {{Power Law}} for {{Fatigue}}, by {{Using Moving Least
  Squares}},} \emph{CMC: Computers, Materials \& Continua}, Vol.~45, No.~1,
  2015, p.~16.

\bibitem[{Sbarufatti et~al.(2014)Sbarufatti, Manes, and
  Giglio}]{sbarufattiApplicationSensorTechnologies2014}
Sbarufatti, C., Manes, A., and Giglio, M., \enquote{Application of Sensor
  Technologies for Local and Distributed Structural Health Monitoring,}
  \emph{Structural Control and Health Monitoring}, Vol.~21, No.~7, 2014, pp.
  1057--1083.
\newblock \doi{10.1002/stc.1632}.

\bibitem[{Wang et~al.(2015)Wang, Haynes, Huang, Dong, and
  Atluri}]{wangUseHighPerformanceFatigue2015}
Wang, H.-K., Haynes, R., Huang, H.-Z., Dong, L., and Atluri, S.~N.,
  \enquote{The {{Use}} of {{High-Performance Fatigue Mechanics}} and the
  {{Extended Kalman}} / {{Particle Filters}}, for {{Diagnostics}} and
  {{Prognostics}} of {{Aircraft Structures}},} \emph{CMES: Computer Modeling in
  Engineering \& Sciences}, Vol. 105, No.~1, 2015, pp. 1--24.

\bibitem[{Henry et~al.(2020)Henry, Cole, Kube, Fudger, Haynes, Mogonye, and
  Weiss}]{henryEvaluationEarlyFatigue2020}
Henry, T.~C., Cole, D.~P., Kube, C.~M., Fudger, S.~J., Haynes, R.~A., Mogonye,
  J.-E., and Weiss, V., \enquote{Evaluation of {{Early Fatigue Signatures}} in
  {{Lightweight Aluminum Alloy}} 7075,} \emph{Experimental Mechanics}, Vol.~60,
  No.~2, 2020, pp. 205--216.
\newblock \doi{10.1007/s11340-019-00547-7}.

\end{thebibliography}

\end{document}